\documentclass[
 reprint,
 amsmath,amssymb,
 aps,
 prb,
 longbibliography
]{revtex4-2}

\usepackage{graphicx}
\usepackage{dcolumn}
\usepackage{bm}
\usepackage{physics}
\usepackage{float}
\usepackage{color}
\usepackage{hyperref}
\hypersetup{
     colorlinks   = true,
     linkcolor    = blue,
     citecolor    = blue,
     urlcolor     = blue
}
\newcommand{\Eq}[1]{Eq.~(\ref{#1})}
\newcommand{\Sec}[1]{Sec.~\ref{#1}}
\newcommand{\Fig}[1]{Fig.~\ref{#1}}

\begin{document}

\title{Magnetic toroidal monopoles from relativistic polarization responses to magnetic field gradients
}

\author{Taisei Yamanaka}
 \email{yamanaka@phys.sci.hokudai.ac.jp}
\author{Takumi Sato}
 \email{sato@phys.sci.hokudai.ac.jp}
\author{Satoru Hayami}
 \email{hayami@phys.sci.hokudai.ac.jp}
\affiliation{
Graduate School of Science, Hokkaido University, Sapporo 060-0810, Japan
}

\date{\today}

\begin{abstract} 
The magnetic toroidal monopole, a time-reversal-odd scalar, has attracted attention through its characteristic responses, such as electric-field-induced nonreciprocal directional dichroism observed in Co$_2$SiO$_4$.
However, its evaluation in crystalline solids remains unresolved, as it cannot be defined within conventional multipole expansions or thermodynamic formulations.
In this paper, we propose a theoretical framework to evaluate the magnetic toroidal monopole in periodic crystals based on the response of relativistic electric polarization to a magnetic field gradient.
By incorporating the magnetic-field-gradient correction to the relativistic polarization, we derive an explicit expression for the magnetic toroidal monopole beyond symmetry arguments.
The resulting expression is formulated in terms of geometric quantity such as Berry curvatures and orbital magnetic moment defined in an extended parameter space spanning momentum, magnetic field, and electric field. 
We further perform model calculations for an antiferromagnetic system hosting a magnetic toroidal monopole and confirm that the proposed quantity is finite.  
These results provide a practical route to characterize magnetic toroidal monopoles in crystalline solids and clarify their quantum geometric nature.
\end{abstract}

\maketitle

\section{Introduction}
The breaking of time-reversal symmetry underlies a wide range of magnetic phenomena. 
A typical example is the anomalous Hall effect originating from finite Berry curvature in momentum space~\cite{PhysRev.95.1154, SMIT195839, PhysRev.160.421, berger1970slide,nozieres1973simple,Loss_PhysRevB.45.13544, ye1999berry,jungwirth2002, nagaosa2010anomalous, gosalbez2015chiral}. 
This effect was long attributed to net magnetization in ferromagnets; however, recent studies have demonstrated that it can also occur in antiferromagnets with broken time-reversal symmetry~\cite{Solovyev_PhysRevB.55.8060, Sivadas_PhysRevLett.117.267203, Chen_PhysRevLett.112.017205, naka2020anomalous, vsmejkal2020crystal, Hayami_PhysRevB.103.L180407, Chen_PhysRevB.106.024421,naka2022anomalous}, so-called altermagnets~\cite{vsmejkal2022beyond}.
In addition, when spatial inversion symmetry is broken, nonreciprocal transport phenomena~\cite{rikken1997observation, wakatsuki2017nonreciprocal, tokura2018nonreciprocal, watanabe2020nonlinear, yatsushiro2022analysis, Hayami_PhysRevB.106.014420, Suzuki_PhysRevB.105.075201, nagaosa2024nonreciprocal} and magnetoelectric effects~\cite{curie1894symetrie, popov1999magnetic, Fiebig0022-3727-38-8-R01, Spaldin2005renaissance, EdererPhysRevB.76.214404, Hayami_PhysRevB.90.081115, thole2018magnetoelectric} can emerge.
These phenomena in antiferromagnets originate from symmetry-breaking electronic degrees of freedom that serve as sources of diverse physical responses.

The electronic states and associated physical phenomena induced by breakings of spatial inversion and time-reversal symmetry can be described in a unified manner using multipole representation theory~\cite{suzuki2018first, kusunose2022generalization, hayami2024unified}. 
Within this framework, time-reversal symmetry breaking is characterized by magnetic multipoles and magnetic toroidal multipoles. 
For example, a magnetic dipole is a source of the anomalous Hall effect, while a magnetic toroidal dipole is a source of nonreciprocal transport and linear magnetoelectric effect. 

\begin{figure}[htbp]
\includegraphics[width=\linewidth]{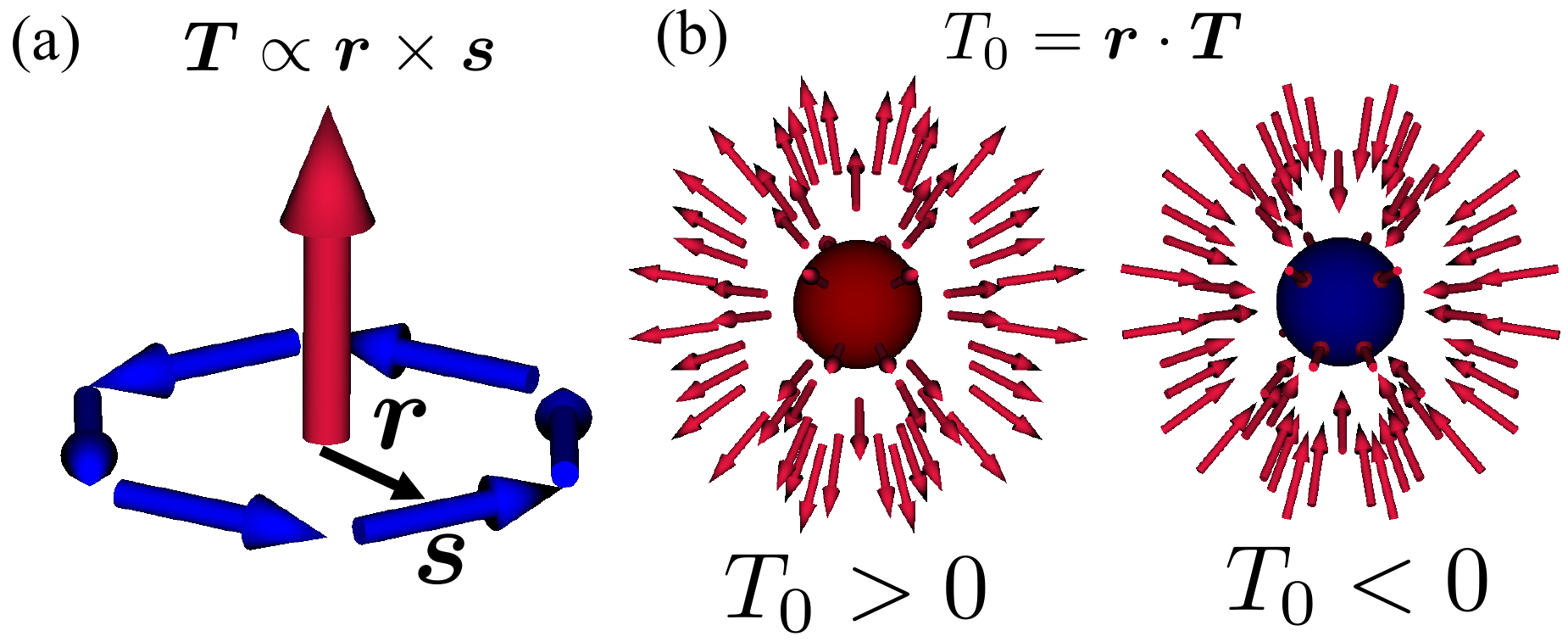}
\caption{
(a) Schematic illustration of the magnetic toroidal dipole. 
The blue arrows indicate the spin directions, while the central arrow denotes the magnetic toroidal dipole. 
(b) Schematic illustration of the MTM. 
The left and right panels correspond to positive and negative MTMs, respectively.
The illustrations are generated using QtDraw~\cite{Kusunose_PhysRevB.107.195118}.
}
\label{fig:t0} 
\end{figure}

The magnetic toroidal monopole (MTM), which is denoted as $T_0$, is a time-reversal-odd scalar corresponding to the source of the magnetic toroidal dipole.
Schematic illustrations of the magnetic toroidal dipole and the MTM are shown in \Fig{fig:t0}.
It is formally expressed as the inner product of the position vector $\vb*{r}$ and the magnetic toroidal dipole $\vb*{T}$,
\begin{align}
  T_0 = \vb*{r} \cdot \vb*{T}
  \label{T01}.
\end{align}
This quantity has been discussed within the cluster-multipole framework~\cite{hayami2023time} and in many-body systems~\cite{kuniyoshi2026theory}.
From a symmetry viewpoint, the MTM can also be 
expressed as coupling between four types of dipoles,
\begin{align}
  T_0 \sim \vb*{Q} \cdot \vb*{T}
  \label{T02}, \\
  T_0 \sim \vb*{G} \cdot \vb*{M}
  \label{T03},
\end{align}
where $\vb*{Q}$, $\vb*{G}$, and $\vb*{M}$ denote electric, electric toroidal, and magnetic dipoles, respectively.
Such couplings imply characteristic responses unique to the MTM, including electric-field-induced spin vortices and magnetic-field-induced rotational distortions~\cite{hayami2023time}. 
Indeed, electric-field-induced nonreciprocal directional dichroism related to the coupling in Eq.~(\ref{T02})~\cite{schmid2001ferrotoroidics, schmid2008some} has been experimentally observed in $\mathrm{Co_2SiO_4}$\cite{hayashida2025electric,kato2026electric}.

However, a method for evaluating the MTM in periodic crystals has not yet been established.
One difficulty arises from the fact that the naive expression $\bm{r}\cdot\bm{T}$ vanishes identically because $\bm{T}\propto \bm{r}\times\bm{s}$~\cite{Spaldin_0953-8984-20-43-434203, hayami2024unified}.
Another difficulty is associated with the position operator $\bm{r}$, which is not well defined for Bloch states in periodic crystals.
Although recent studies have shown that such difficulties can be overcome within a thermodynamic formulation~\cite{shi2007quantum,gao2018microscopic,shitade2018theory,gao2018orbital,shitade2019theory,daido2020thermodynamic,oike2025thermodynamic,sato2026,sato2026orbitalmagneticoctupolecrystalline}, 
this approach is not directly applicable to the MTM. 
For example, a magnetic octupole proportional to $s_i r_j r_k$ ($i,j,k=x,y,z$) can be defined within the thermodynamic formalism through its coupling to the second spatial derivative of the magnetic field, $\partial_j \partial_k B_i$~\cite{oike2025thermodynamic,sato2026}. 
In contrast, there is no corresponding thermodynamic quantity for the MTM, which is associated with the antisymmetric part of $r_j r_k$.    

In this study, we develop a theoretical framework for describing the MTM in crystalline solids from the viewpoint of response theory. 
By examining the response of relativistic electric polarization~\cite{katsura2005spin,hayami2024analysis} to a magnetic field gradient, we identify the contribution associated with the MTM in a gauge-invariant manner, thereby providing a route to characterize it in periodic systems. 
The formulation naturally leads to expressions involving quantum geometric quantities defined in an extended parameter space, reflecting the interplay between momentum, magnetic field, and electric field.
Furthermore, we perform model calculations for an antiferromagnetic system hosting the MTM and confirm that the proposed framework yields a finite contribution, demonstrating its applicability to realistic systems.

The remainder of this paper is organized as follows: 
In \Sec{Methods}, we derive the response tensor of relativistic electric polarization to a magnetic field gradient. 
In \Sec{Analysis}, we discuss the physical meaning of the obtained expression and its relation to thermodynamic multipoles. 
In \Sec{Model}, we introduce a minimal antiferromagnetic model exhibiting the MTM and present numerical results based on the derived formula. 
Finally, in \Sec{Summary}, we summarize this paper.

Throughout this paper, we use the units of $k_{\text{B}} = c = \hbar = 1$, where $k_{\text{B}}$ is the Boltzmann constant and $c$ is the speed of light.

\section{Response theory of relativistic polarization to magnetic field gradients}\label{Methods}

We first discuss a general viewpoint for characterizing the MTM. 
Since the MTM is a rank-0 scalar quantity, it is natural to describe it in terms of higher-rank tensors and their contractions. 
In particular, a rank-3 tensor, such as the magnetic octupole, provides a suitable starting point, as it shares the same spatial inversion and time-reversal parities as the MTM.
The conventional magnetic octupole is proportional to $s_i r_j r_k$, where the factor $r_j r_k$ is constructed from a single polar vector.
As a result, it is symmetric 
under the interchange of $r_j$ and  $r_k$, which prevents the formation of a rank-0 scalar component.
To overcome this limitation, we reinterpret the product $r_j r_k$ as a combination of two independent polar vectors, $r_j R_k$. 
This enables the construction of a general rank-3 tensor, $s_i r_j R_k$, which includes the completely antisymmetric component; contracting it with the Levi-Civita tensor, $\epsilon_{ijk} s_i r_j R_k$, yields a time-reversal-odd rank-0 scalar that can be identified with $T_0$.
Since symmetry alone does not uniquely determine a bulk MTM, we adopt a response-based definition using the completely antisymmetric component of the relativistic electric-polarization response to a magnetic-field gradient.
The resulting quantity exhibits the expected MTM-related scalar couplings and becomes finite when the MTM is symmetry-allowed.

We begin with the thermodynamic relation for the standard magnetic octupole $M_{ijk}$ at zero temperature in insulators~\cite{oike2025thermodynamic,sato2026,shitade2025intrinsic}.
\begin{align}
\label{eq: thermo_relation}
  \frac{\partial M_{ijk}}{\partial \mu} = -e\frac{\partial P_k}{\partial (\partial_j B_i)}.
\end{align}
Here, $\mu$, $e$, and $\vb*{P}$ denote the chemical potential, the electron charge, and the electric polarization, respectively.
This relation indicates that the chemical-potential derivative of the magnetic octupole is directly related to the electric polarization induced by a magnetic field gradient.
By definition, $M_{ijk}$ is symmetric under the interchange of $j$ and $k$ and thus has no completely antisymmetric component. 

Although the polarization $\vb*{P}$ in Eq.~(\ref{eq: thermo_relation}) usually refers to the charge-induced polarization, it is modified by the spin degree of freedom at the relativistic level. 
The corresponding relative correction can be expressed in terms of spin currents~\cite{katsura2005spin,hayami2024analysis}, which are given by 
\begin{align}
  \tilde{\vb*{P}}^{\rm (s)} = \xi \vb*{s} \times \vb*{v} = \xi \tilde{\vb*{P}}, 
\end{align}
where $\xi$ is the relativistic coefficient $\xi = e / 2mc^2$ and $\tilde{\vb*{P}} = \vb*{s} \times \vb*{v}$.
Here, $\vb*{s}$ and $\vb*{v}$ represent the spin and velocity operators, respectively.  
By introducing $\tilde{\vb*{P}}$, one can obtain the completely antisymmetric component of $M_{ijk}$ in Eq.~(\ref{eq: thermo_relation}) in a gauge-invariant form. 

Building on this formulation, we evaluate the MTM from the completely antisymmetric component of the response tensor describing the variation of $\tilde{\vb*{P}}$ under a magnetic field gradient. 
Specifically, we define
\begin{align}\label{eq:dif}
  \chi_{ijk} = \xi \frac{\partial \tilde{P}_k}{\partial (\partial_j B_i)}, \hspace{3mm} \chi^{T_0} = \frac{1}{3!}\epsilon_{ijk} \chi_{ijk},
\end{align}
where 
$\chi_{ijk}$ is a general rank-3 tensor and $\chi^{T_0}$ represents the MTM contribution.
This equation is valid for both metals and insulators at any temperature.

The completely antisymmetric component in Eq.~(\ref{eq:dif}) is directly related to the antisymmetric part of the magnetic-field gradient.
In particular, contraction with the Levi-Civita tensor gives a contribution proportional to $\nabla\times\vb*{B}$.
Therefore, $\chi^{T_0}$ characterizes the relativistic electric polarization induced by a magnetic-field configuration with a finite curl.
Experimentally, such a perturbation could in principle be realized either by applying a controlled spatially nonuniform magnetic field with a finite antisymmetric gradient~\cite{yamanaka2025nonlinear} or by driving an external current that generates a finite $\nabla\times\vb*{B}$, as implied by Amp\`ere's law in the static limit.
The corresponding MTM-related response can then be probed through the induced relativistic electric polarization.

To evaluate $\chi_{ijk}$, we employ the Kubo formalism and calculate the polarization-magnetization correlation function as
\begin{widetext}
  \begin{align}\label{eq:correlation}
  \chi_{\tilde{P}_k, M_i}(\vb*{q}, \omega) = 
  -g\mu_{\text{B}}\xi\sum_{n,m}^{} \int \frac{d^dk}{(2\pi)^d} \bra{n\vb*{k}_-} \frac{1}{2}\left(\hat{\tilde{r}}_k (\vb*{k}_+) + \hat{\tilde{r}}_k (\vb*{k}_-) \right) \ket{m\vb*{k}_+}\bra{m\vb*{k}_+} \hat{s}_i \ket{n\vb*{k}_-}
  \frac{f_{n\vb*{k}_-} - f_{m\vb*{k}_+}}{\epsilon_{n\vb*{k}_-} - \epsilon_{m\vb*{k}_+} + \omega + i\delta},
\end{align}
\end{widetext}
which describes the linear response $\delta\tilde{P}_k(\vb*{q}, \omega) =  \chi_{\tilde{P}_k, M_i}(\vb*{q}, \omega)B_i(\vb*{q}, \omega)$ with the wave vector $\bm{q}$ and the frequency $\omega$.
$\tilde{P}_k$ corresponds to the expectation value of polarization operator; $\tilde{P}_k = \ev{\hat{\tilde{r}}_k}$ with $\vb*{\tilde{r}}=\vb*{s} \times \vb*{v}$. 
Here, $g$, $\mu_{\text{B}}$, $\hat{s}_i$, and $d$ represent the spin g factor, Bohr magneton, spin operator, and spatial dimension, respectively.
In addition, we adopt the following notation;
\begin{align}
  \hat{\mathcal{H}}_{\vb*{k}}\ket{n\vb*{k}} &= \epsilon_{n\vb*{k}}\ket{n\vb*{k}}, \hspace{3mm} f_{n\vb*{k}} = (1 + e^{\beta(\epsilon_{n\vb*{k}} - \mu)})^{-1}, \nonumber\\
  \vb*{k}_\pm &= \vb*{k}\pm\frac{\vb*{q}}{2}, \hspace{3mm} \hat{\tilde{r}}_k(\vb*{k}) = \frac{1}{2}\epsilon_{kab}\left\{ \hat{v}_a(\vb*{k}), \hat{s}_b\right\}_+,
\end{align}
where $\hat{\mathcal{H}}_{\vb*{k}}$ is the Bloch Hamiltonian with eigenvalues $\epsilon_{n\bm{k}}$ for eigenstates $\ket{n\bm{k}}$ ($n$ labels the band index and $\vb*{k}$ is the crystal momentum),  $f_{n\vb*{k}}$ is the Fermi distribution function, and $\hat{v}_{j}(\vb*{k})$ is the velocity operator; $\hat{v}_{j}(\vb*{k}) = \partial_{k_j} \hat{\mathcal{H}}_{\vb*{k}}$ ($\partial_{k_j}= \partial / \partial k_{j}$).

Taking the static limit $\omega \to 0$ and expanding $\chi_{\tilde{P}_k, M_i}(\vb*{q}, \omega)$ to the first order in $\vb*{q}$, we obtain $\chi_{ijk}$ as
  \begin{align}\label{eq:chi}
  \chi_{ijk} = \lim_{\vb*{q} \rightarrow 0} -i\partial_{q_j}\lim_{\delta \rightarrow 0} \chi_{\tilde{P}_k, M_i}(\vb*{q}, 0).
\end{align}
Carrying out the calculations, the response tensor is expressed in the gauge-invariant form as
\begin{widetext}
\begin{align}\label{eq:responce}
  \chi_{ijk} 
  = -g\mu_{\text{B}}\xi\int_{}^{}\frac{d^dk}{(2\pi)^d} &\sum_{n}^{} \sum_{m}^{\neq n}
  \Bigg[ 
-\frac{1}{2}\Bigg\{(s^i_n + s^i_m) \Omega^{v_j, \tilde{r}_k}_{nm} + (v^j_n + v^j_m) \Omega^{\tilde{r}_k, s_i}_{nm} + (\tilde{r}^k_n + \tilde{r}^k_m) \Omega^{s_i, v_j}_{nm} \Bigg\}f_n \nonumber\\
&- \Bigg\{s^i_n m^{v_j, \tilde{r}_k}_{nm} + v^j_n m^{\tilde{r}_k, s_i}_{nm}+ \tilde{r}^k_n m^{s_i, v_j}_{nm} \Bigg\}f'_n + \sum_{l}^{\neq n, m}X^{ijk}_{nml}f_n
  \Bigg],
\end{align}
where
\begin{align}
  \Omega^{A, B}_{nm} &= -2\Im\left[ \frac{A_{nm}B_{mn}}{\epsilon^2_{nm}} \right]
  \label{eq:def1} , \\
  m^{A, B}_{nm} &= \Im\left[ \frac{A_{nm}B_{mn}}{\epsilon_{nm}}  \right]
  \label{eq:def2} , \\
      X^{ijk}_{nml} &= \Im\Bigg[ \frac{s^i_{ln}v^j_{ml}\tilde{r}^k_{nm} - s^i_{nm}v^j_{ml}\tilde{r}^k_{ln}}{\epsilon_{nm}\epsilon_{ml}} + \frac{s^i_{ml}v^j_{ln}\tilde{r}^k_{nm} - s^i_{nm}v^j_{ln}\tilde{r}^k_{ml}}{\epsilon_{nm}\epsilon_{ln}} \Bigg] \label{eq:def3} .
\end{align}
\end{widetext}
We use the abbreviations $A^i_{nm} = \bra{n\vb*{k}} \hat{A}_i \ket{m\vb*{k}}$, $A^i_n = A^i_{nn}$ with $\hat{\vb*{A}} = \hat{\vb*{s}}, \hat{\vb*{v}}, \hat{\tilde{\vb*{r}}}$, $\hat{\vb*{v}} = \hat{\vb*{v}}(\vb*{k}), \hat{\tilde{\vb*{r}}} = \hat{\tilde{\vb*{r}}}(\vb*{k})$, $\epsilon_n = \epsilon_{n\vb*{k}}$, $\epsilon_{nm} = \epsilon_n  - \epsilon_m$, $f_n = f_{n\vb*{k}}$ and $f'_n = \partial f_{n} / \partial \epsilon_{n}$.
The detailed derivation is presented in Appendix~\ref{derivation}.
We also show that the expression remains valid even in the presence of band touchings, including degenerate points in Appendix~\ref{degeneracy}.

The response tensor in Eq.~(\ref{eq:responce}) consists of several contributions involving $\Omega^{A,B}_{nm}$, $m^{A,B}_{nm}$, and $X^{ijk}_{nml}$, which encode interband processes mediated by the operators $\hat{v}$, $\hat{s}$, and $\hat{\tilde{r}}$. 
The first two quantities, $\Omega^{A,B}_{nm}$ and $m^{A,B}_{nm}$, are generalizations of well-known geometric quantities in periodic crystals. 
For example, by taking $A=v^j$ and $B=v^k$, $\Omega^{A,B}_{nm}$ and $m^{A,B}_{nm}$ correspond to the band-resolved Berry curvature and orbital magnetic moment, respectively, where the Berry curvature and orbital magnetic moment obtained within semiclassical theory~\cite{chang1995berry,sundaram1999wave,PhysRevB.53.7010,RevModPhys.82.1959}, 
are given by
\begin{align}
  \Omega^{jk}_n &= \sum_{m}^{m \neq n}-2\Im\left[ \frac{v^j_{nm}v^k_{mn}}{\epsilon^2_{nm}} \right], \\
  m^{jk}_n &= \sum_{m}^{m \neq n}\Im\left[ \frac{v^j_{nm}v^k_{mn}}{\epsilon_{nm}} \right] .
\end{align}

Since the velocity operator is given by the momentum derivative of the Hamiltonian, $\hat{v}_j = \partial_{k_j}\hat{\mathcal{H}}_{\vb*{k}}$, these quantities can be expressed in terms of geometric tensors in momentum ($k$-$k$) space. 
They are related to the quantum geometric tensor of the $n$th band, 
\begin{align}
  T^{jk}_n &= \bra{\partial_{k_j}n\vb*{k}} \left(1 - \ket{n}\bra{n} \right)\ket{\partial_{k_k}n\vb*{k}}, \\
  \tilde{T}^{jk}_n &= \bra{\partial_{k_{j}}n\vb*{k}} \left(\hat{\mathcal{H}}_{\vb*{k}} - \epsilon_{n\vb*{k}} \right)\ket{\partial_{k_{k}}n\vb*{k}}, 
\end{align}
which satisfy $\Omega^{jk}_n = -2\mathrm{Im}T^{jk}_n$ and $m^{jk}_n = -\mathrm{Im}\tilde{T}^{jk}_n$~\cite{ma2010abelian,kang2025measurements}. 
In the present formulation, the operators $\hat{\vb*{s}}$ and $\hat{\tilde{\vb*{r}}}$ can also be regarded as derivatives of a parameter-dependent Hamiltonian,
\begin{align}
\hat{\mathcal{H}} = \hat{\mathcal{H}}_{\vb*{k}} + \vb*{h}\cdot\hat{\vb*{s}} + \vb*{e}\cdot\hat{\tilde{\vb*{r}}},
\end{align}
with $\vb*{h} = -g\mu_{\mathrm{B}}\vb*{B}$ and $\vb*{e} = \xi\vb*{E}$, such that $\partial_{h_i}\hat{\mathcal{H}} = \hat{s}_i$ and $\partial_{e_i}\hat{\mathcal{H}} = \hat{\tilde{r}}_i$.
From this viewpoint, $\Omega^{A,B}_{nm}$ and $m^{A,B}_{nm}$ in Eq.~(\ref{eq:responce}) can be interpreted as geometric quantities defined in an extended parameter space spanned by crystal momentum ($k$), magnetic field ($h$), and electric field ($e$). 
This highlights that the MTM-related response function is intrinsically linked to the quantum geometry of electronic states beyond conventional momentum-space formulations.

\section{Bulk formulation of magnetic toroidal monopole}\label{Analysis}
Combining Eqs.~(\ref{eq: thermo_relation}) and (\ref{eq:dif}), the MTM in crystalline solids can be defined through the antisymmetric component of the response tensor. 
The corresponding relation between the MTM and the linear response is given by
\begin{align}
  \frac{\partial \tilde{M}_{ijk}}{\partial \mu} := \xi \frac{\partial \tilde{P}_k}{\partial (\partial_j B_i)},
\end{align} 
where $\tilde{M}_{ijk}$ denotes a generalized rank-3 tensor that includes the completely antisymmetric component.
Since the chemical potential dependence enters only through the Fermi distribution function, the integration can be carried out straightforwardly as 
  \begin{align}
    \tilde{M}_{ijk} &= \int_{-\infty}^{\mu}\xi \frac{\partial \tilde{P}_k}{\partial (\partial_j B_i)} d\mu' .
  \end{align}
The lower limit $\mu=-\infty$ corresponds to the empty-system limit.
With this convention, $\tilde{M}_{ijk}$ vanishes at $\mu=-\infty$, and hence so does $\widetilde{T}_0$, thereby fixing the additive integration constant.
The resulting quantity is invariant under a uniform shift of the energy origin.
Carrying out the integration yields
\begin{widetext}
\begin{align}\label{eq:MO}
  \tilde{M}_{ijk} = -g\mu_{\text{B}}\int_{}^{}\frac{d^dk}{(2\pi)^d} &\sum_{n}^{} \sum_{m}^{\neq n}
  \Bigg[
\frac{1}{2}\Bigg\{(s^i_n + s^i_m) \Omega^{v_j, \tilde{r}_k}_{nm} + (v^j_n + v^j_m) \Omega^{\tilde{r}_k, s_i}_{nm} + (\tilde{r}^k_n + \tilde{r}^k_m) \Omega^{s_i, v_j}_{nm} \Bigg\}\mathcal{G}_n \nonumber\\
&+ \Bigg\{s^i_n m^{v_j, \tilde{r}_k}_{nm} + v^j_n m^{\tilde{r}_k, s_i}_{nm}+ \tilde{r}^k_n m^{s_i, v_j}_{nm} \Bigg\}f_n - \sum_{l}^{\neq n, m}X^{ijk}_{nml}\mathcal{G}_n
  \Bigg].
\end{align}
\end{widetext}
Here, we use $\int_{-\infty}^{\mu} f_n d\mu' = -\mathcal{G}_n$, $\int_{-\infty}^{\mu} f'_n d\mu' = -f_n$.
$\mathcal{G}_n  = -T\log\left\{1 + e^{-\left(\epsilon_{n\vb*{k}} - \mu\right)/T}\right\}$ denotes the grand potential density and should be distinguished from the electric toroidal dipole $\vb*{G}$.
In \Eq{eq:MO}, the terms involving the off-diagonal components of $\hat{\tilde{\vb*{r}}}$ reduce to those of the thermodynamic spin magnetic octupole~\cite{oike2025thermodynamic,sato2026} when $\hat{\tilde{\vb*{r}}}$ is replaced by the position operator in momentum space, $\hat{\tilde{\vb*{r}}}\to -i\nabla_{\vb*{k}}$.

The MTM in periodic crystals is then defined as the completely antisymmetric component, $\tilde{T}_0 := \frac{1}{3!}\epsilon_{ijk} \tilde{M}_{ijk}$, which is given by
\begin{align}
  \tilde{T}_0
  &=  \int_{}^{}\frac{d^dk}{(2\pi)^d} \sum_{n}^{} \frac{1}{3!}\bigg[
    \vb*{\tilde{r}}_n \cdot \vb*{T}_n + \vb*{v}_n \cdot \vb*{Q}_n + \vb*{s}_n \cdot \vb*{G}_n + F_n
  \bigg] .
\end{align}
Here, 
\begin{align}
   T^k_n &= -g\mu_{\text{B}} \epsilon_{ijk}\left\{ \Omega^{s_i, v_j}_n \mathcal{G}_n + m^{s_i, v_j}_nf_n\right\}, \\
   Q^k_n &= -g\mu_{\text{B}} \epsilon_{ijk}\left\{ \Omega^{\tilde{r}_i, s_j}_n \mathcal{G}_n + m^{\tilde{r}_i, s_j}_nf_n\right\}, \\
   G^k_n &= -g\mu_{\text{B}} \epsilon_{ijk}\left\{ \Omega^{v_i, \tilde{r}_j}_n \mathcal{G}_n + m^{v_i, \tilde{r}_j}_nf_n\right\}, \\
   F_n &= -g\mu_{\text{B}} \epsilon_{ijk}
\sum_{m}^{\neq n}\frac{-1}{2}\Bigg\{(s^i_n - s^i_m) \Omega^{v_j, \tilde{r}_k}_{nm} + (v^j_n - v^j_m) \Omega^{\tilde{r}_k, s_i}_{nm} \nonumber\\
&+ (\tilde{r}^k_n - \tilde{r}^k_m) \Omega^{s_i, v_j}_{nm} + 2\sum_{l}^{\neq n, m}X^{ijk}_{nml}\Bigg\}\mathcal{G}_n, 
\end{align}
with $\Omega^{A, B}_{n} = \sum_{m \neq n}^{}\Omega^{A, B}_{nm}$.
This expression indicates that the MTM is described as a linear combination of inner products of vector quantities, $\vb*{\tilde{r}}_n \cdot \vb*{T}_n$, $\vb*{v}_n \cdot \vb*{Q}_n$, and $\vb*{s}_n \cdot \vb*{G}_n$.  
Each of these terms forms a time-reversal-odd scalar, consistent with the symmetry of the MTM: 
$\vb*{\tilde{r}}_n$, $\vb*{Q}_n$, and $\vb*{G}_n$ are even under time reversal, whereas $\vb*{T}_n$, $\vb*{v}_n$, and $\vb*{s}_n$ are odd; similarly, $\vb*{s}_n$ and $\vb*{G}_n$ are even under spatial inversion, while $\vb*{v}_n$, $\vb*{\tilde{r}}_n$, $\vb*{Q}_n$, and $\vb*{T}_n$ are odd.
Similar to Eq.~(\ref{eq:dif}), this expression is gauge-invariant, and remains valid for both metals and insulators at all temperatures.

The quantity $\vb*{T} = \int \frac{d^dk}{(2\pi)^d} \sum_{n} \vb*{T}_n$ corresponds to the thermodynamic spin magnetic toroidal dipole~\cite{gao2018microscopic,shitade2019theory}, which characterizes a vortex-like distribution of magnetic moments in periodic crystals.
The thermodynamic magnetic toroidal dipole is associated with geometric quantities involving the Berry curvature and the orbital magnetic moment in the $h$-$k$ parameter space, reflecting its origin of magnetic-field-induced electric polarization and vice versa. 
In this context, the term $\vb*{\tilde{r}}_n \cdot \vb*{T}_n$, as introduced in Eq.~(\ref{T01}), can be interpreted as a coupling between the relativistic electric polarization and the magnetic toroidal dipole moment. 
This term provides a microscopic realization of the scalar quantity formed by contracting a polar vector with a magnetic toroidal moment, consistent with the intuitive expression for the source of the magnetic toroidal dipole given in Eq.~(\ref{T01}).

Furthermore, from a symmetry viewpoint, $\vb*{Q}_n$ and $\vb*{G}_n$ correspond to the electric dipole and electric toroidal dipole in periodic crystals, respectively.
The electric dipole is associated with the Berry curvature in the $e$-$h$ parameter space, which can be viewed as related to time-dependent-magnetic-field-induced spin currents, whereas the electric toroidal dipole is associated with the Berry curvature in the $k$-$e$ space, related to electric-field-induced spin currents~\cite{Hayami_doi:10.7566/JPSJ.91.113702}.
Accordingly, the terms $\vb*{v}_n \cdot \vb*{Q}_n$ and $\vb*{s}_n \cdot \vb*{G}_n$ represent couplings between fundamental electronic degrees of freedom ($\vb*{v}_n$ and $\vb*{s}_n$) and the dipolar moments.
Although these contributions are consistent with the symmetry-allowed couplings in Eqs.~(\ref{T02}) and (\ref{T03}), our results go beyond this symmetry argument by showing that they originate from microscopic couplings and directly contribute to the MTM, thereby establishing its composite nature in terms of multiple dipolar components. 
In this way, the MTM is composed of intertwined contributions from relativistic polarization, velocity, and spin degrees of freedom, which cannot be defined within the conventional thermodynamic framework~\cite{oike2025thermodynamic, sato2026}.

It is worth noting that the presence of the additional term $F_n$ indicates that the MTM cannot be fully reduced to simple dipole-dipole couplings.
Instead, it involves intrinsic multiband and interband contributions encoded in $X^{ijk}_{nml}$, reflecting the underlying quantum geometry of the electronic structure.

\section{Model calculation}\label{Model}

\begin{figure}[htbp]
\includegraphics[width=\linewidth]{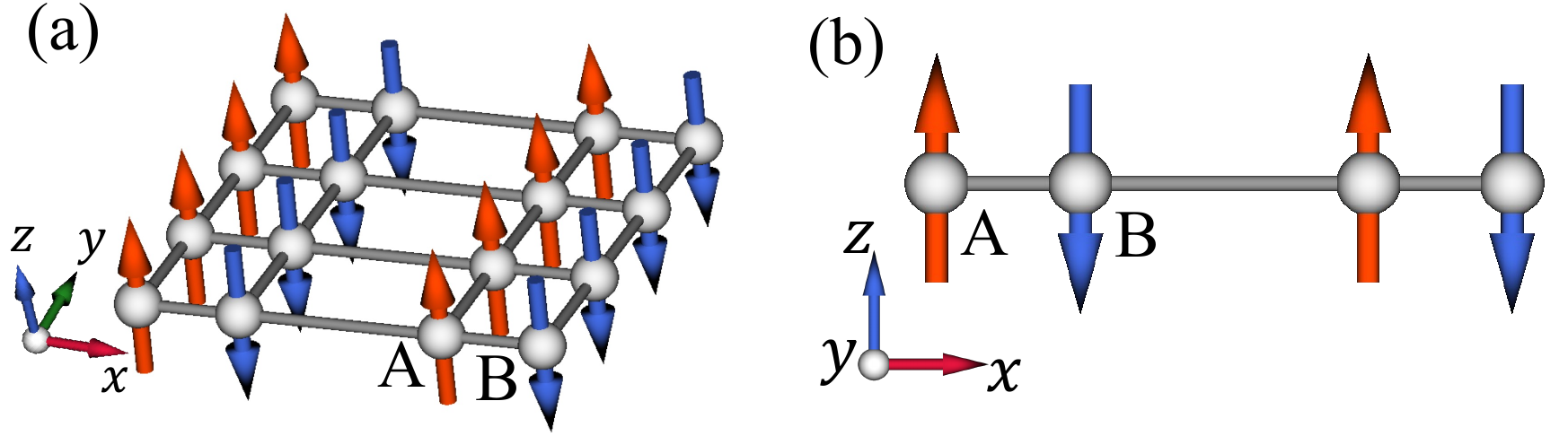}
\caption{
(a,b) Schematic illustrations of the model with magnetic point group $m2m$, shown from different viewing angles. 
The red and blue arrows represent up and down spins, respectively. 
The figures are generated using QtDraw~\cite{Kusunose_PhysRevB.107.195118}.
}
\label{fig:1} 
\end{figure}

  To examine the relationship between the obtained MTM and the internal degrees of freedom of a material, we perform 
  model calculations based on a minimal two-sublattice system shown in \Fig{fig:1}(a) and \Fig{fig:1}(b), which belongs to the magnetic point group $m2m$.
  The model Hamiltonian is given by 
\begin{align}\label{eq:Hamiltonian}
  \mathcal{H} &= \mathcal{H}_{\text{hop}} + \mathcal{H}_{\text{MF}} + \mathcal{H}_{\text{SOI}}, \\
  \mathcal{H}_{\text{hop}} &= \varepsilon_0(\vb*{k}) + \varepsilon_x(\vb*{k})\tau_x + \varepsilon_y(\vb*{k})\tau_y, \\ 
  \mathcal{H}_{\text{MF}} &= -h \tau_z \otimes \sigma_z, \\
  \mathcal{H}_{\text{SOI}} &= \alpha \sin(k_z) \sigma_x,
\end{align}
with 
\begin{align}
    \varepsilon_0(\vb*{k}) &= -2t_y\cos(k_y) -2t_z\cos(k_z), \nonumber\\
    \varepsilon_x(\vb*{k}) &=  -2t_x\cos(k_x), \hspace{3mm} \varepsilon_y(\vb*{k}) = -2t'_x\sin(k_x) .
\end{align}
Here, $\sigma_i$ and $\tau_i$ are the Pauli matrices acting on the spin and sublattice degrees of freedom, respectively.
$\mathcal{H}_{\text{hop}}$ is the hopping Hamiltonian including inter-sublattice hoppings, $t_x$ and $t'_x$, along the $x$ direction and the intra-sublattice hopping along the $y$ direction, $t_y$, and that along the $z$ direction, $t_z$.
We take the lattice constant as the unit of length and set the hopping parameters as 
$t_x = 0.9$, $t'_x = 0.1$, $t_y = t_z = 1$.
$\mathcal{H}_{\text{MF}}$ denotes the mean-field term to describe the collinear antiferromagnetic ordering.
$\mathcal{H}_{\text{SOI}}$ denotes the antisymmetric spin--orbit interaction term by supposing the polar field along the $y$ direction.

From the symmetry viewpoint, each term in the Hamiltonian plays a distinct role in breaking the symmetry. 
The hopping Hamiltonian $\mathcal{H}_{\text{hop}}$ preserves the magnetic point group symmetry $mmm1'$.
When the polar spin--orbit interaction term $\mathcal{H}_{\text{SOI}}$ is introduced, the symmetry is lowered to $m2m1'$ due to the breaking of spatial inversion and one of the mirror symmetries.
Furthermore, the inclusion of the antiferromagnetic mean-field term $\mathcal{H}_{\text{MF}}$ breaks time-reversal symmetry, reducing the symmetry to $m2m$.
Within this symmetry setting, $\mathcal{H}_{\text{MF}}$ gives rise to a magnetic toroidal dipole~\cite{hayami2022nonlinear}, while $\mathcal{H}_{\text{SOI}}$ induces an electric dipole.
Since the MTM can be expressed as the inner product of a magnetic toroidal dipole and an electric dipole [Eq.~(\ref{T02})], the present model provides a minimal and symmetry-consistent platform for realizing the MTM.
In this sense, the MTM emerges as a consequence of the symmetry lowering by $\mathcal{H}_{\text{SOI}}$ and $\mathcal{H}_{\text{MF}}$.

\begin{figure}[htbp]
\includegraphics[width=\linewidth]{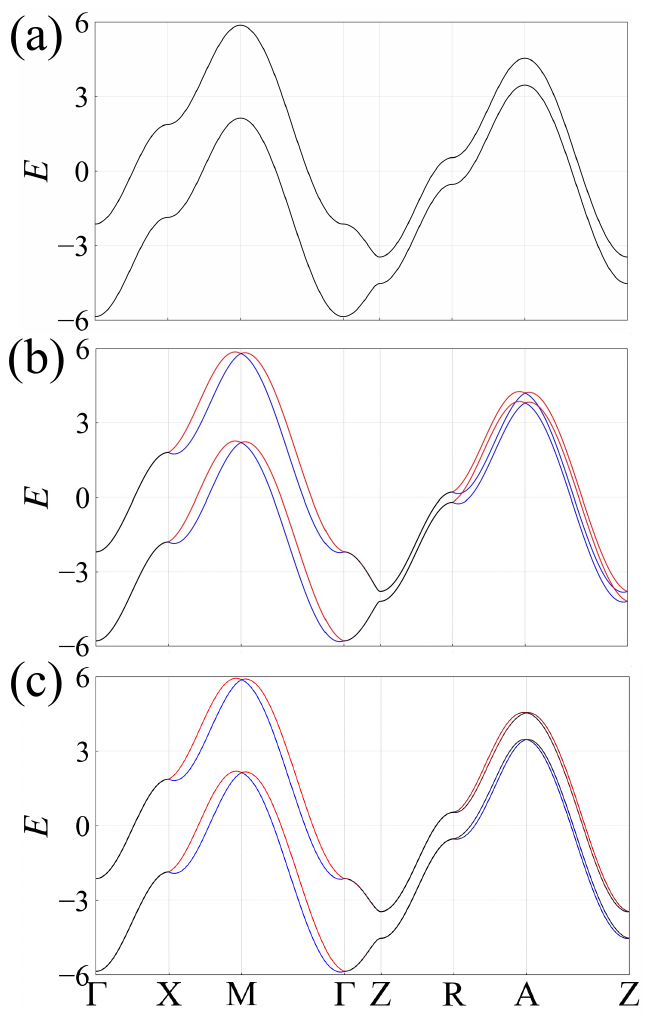}
\caption{
(a)--(c) Band structures calculated from Eq.~(\ref{eq:Hamiltonian}) for (a) $(h,\alpha) = (0.5, 0)$,(b) $(0, 0.5)$, and (c) $(0.5, 0.5)$, respectively.
The high-symmetry points in the Brillouin zone are defined by the following fractional coordinates (in units of the reciprocal lattice vectors): $\Gamma = (0, 0, 0), \mathrm{X} = (0.5, 0, 0), \mathrm{M} = (0.5, 0.5, 0), \mathrm{Z} = (0, 0, 0.5), \mathrm{R} = (0, 0.5, 0.5),$ and $\mathrm{A} = (0.5, 0.5, 0.5)$.
The red and blue colors in (b) indicate the positive and negative spin polarizations along the $x$ direction.
}
\label{fig:2} 
\end{figure}

Figures~\ref{fig:2}(a) and (b) show the band structures for $h = 0.5$ and $\alpha = 0$ 
and $h = 0$ and $\alpha = 0.5$, respectively. 
In \Fig{fig:2}(b), the antisymmetric spin splitting appears along the X--$\Gamma$ and R--Z directions, reflecting the effect of the antisymmetric spin--orbit interaction $\mathcal{H}_{\text{SOI}}$.
Figure~\ref{fig:2}(c) shows the band structure for $(h,\alpha)=(0.5,0.5)$, illustrating the combined effects of the antiferromagnetic mean field and the antisymmetric spin--orbit interaction.

\begin{figure}[htbp]
  \centering
\includegraphics[width=\linewidth]{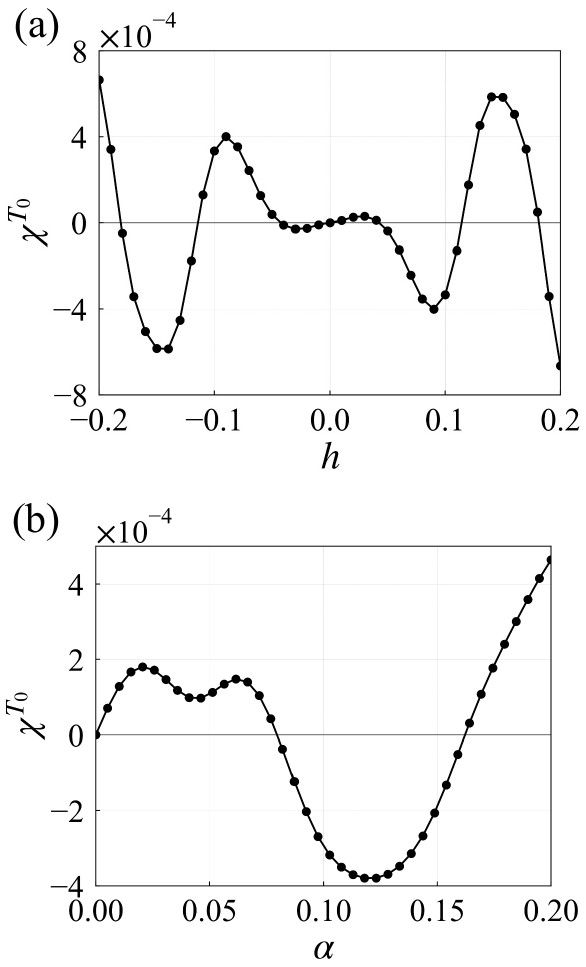}
\caption{
Dependence of $\chi^{T_0}$ on (a) $h$ and (b) $\alpha$ for fixed $\alpha = 0.1$ and $h = 0.1$, respectively.
The parameters are set to $\mu = 1$ and $T = 0.01$.
}
\label{fig:3} 
\end{figure}

Figure~\ref{fig:3}(a) [Figure~\ref{fig:3}(b)] shows the $h$ ($\alpha)$ dependence of $\chi^{T_0}$ at fixed $\alpha = 0.1$ ($h = 0.1$).
The number of unit cells is set to $256^3$.
As expected from the preceding symmetry argument, we find that $\chi^{T}_{0}$ vanishes when either $\alpha$ in the $\mathcal{H}_{\text{SOI}}$ or $h$ in the $\mathcal{H}_{\text{MF}}$ is zero. 
Moreover, the numerical results confirm the expected odd parity with respect to both symmetry-breaking parameters,
$\chi^{T_0}(-h,\alpha)=-\chi^{T_0}(h,\alpha)$ and $\chi^{T_0}(h,-\alpha)=-\chi^{T_0}(h,\alpha)$.

In the present parameter regime, where the chemical potential lies in the metallic regime, the response is predominated by the Fermi-surface contribution proportional to $f'_n$. 
Consequently, $\chi^{T_0}$ is highly sensitive to the evolution of the Fermi surface induced by changes in $h$ and $\alpha$.
This sensitivity leads to a nonmonotonic dependence of $\chi^{T_0}$, which can be attributed to band crossings and the redistribution of spectral weight near the Fermi level. 
It is noted that the behavior of response functions associated with multipoles depends on the specific model, and sign reversals due to variations in model parameters are common.
Similar behavior has also been observed in other multipole systems. 
Furthermore, it has been demonstrated that multipoles contribute to the response even at chemical potentials where the thermodynamic multipoles become zero~\cite{sato2026conductivity}.
\begin{figure*}[htbp]
\includegraphics[width=\linewidth]{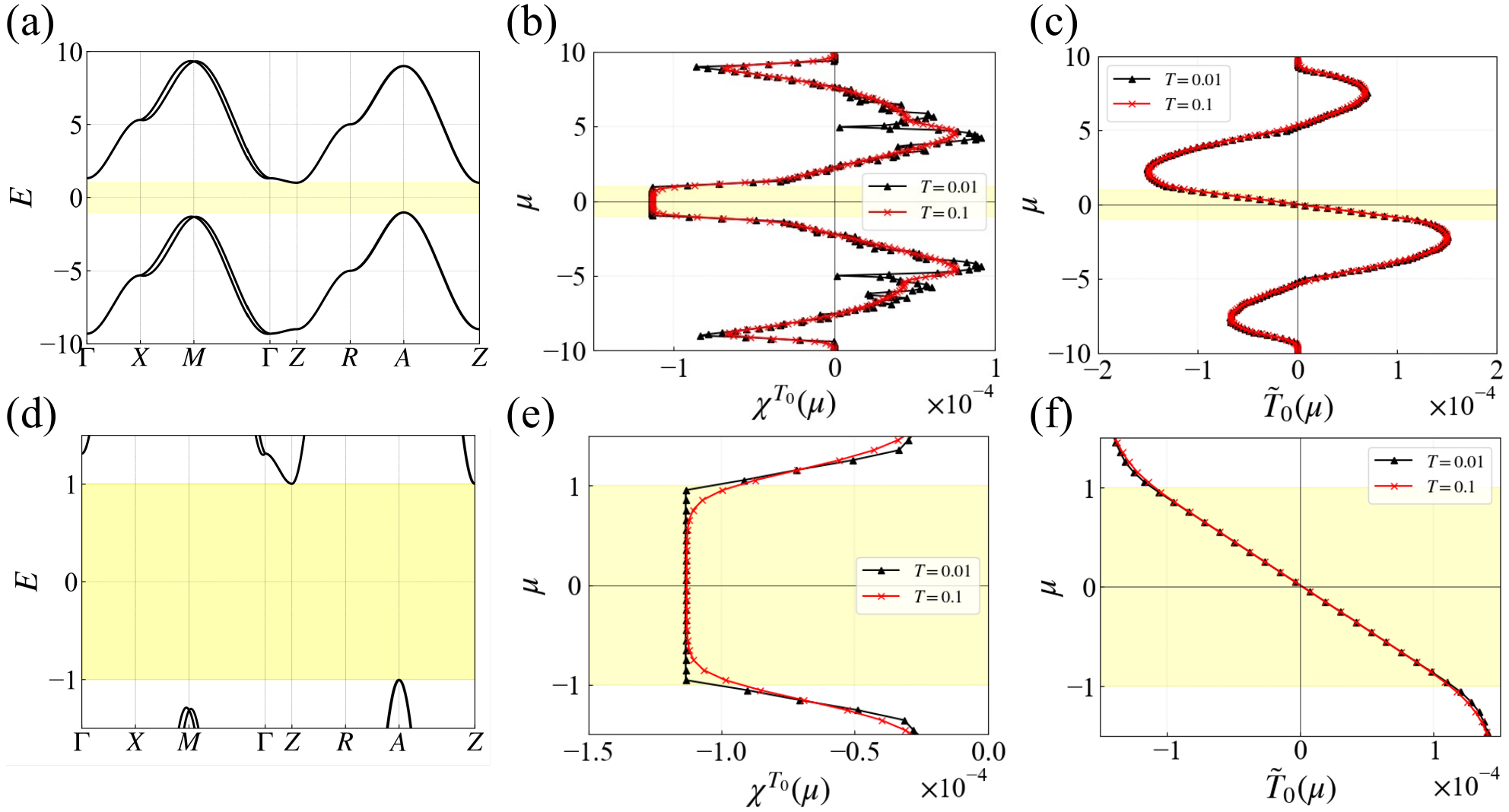}
\caption{
(a), (d) Band structures calculated from Eq.~(\ref{eq:Hamiltonian}).
(b), (e) Chemical potential dependence $\mu$ of $\chi^{T_0}$.
(c), (f) Chemical potential dependence $\mu$ of $\tilde{T}_0$.
In all panels, the parameters are set to $h = 5$ and $\alpha = 1$. 
Panels (d)--(f) show enlarged views of panels (a)--(c) around $-1 \lesssim \mu \lesssim 1$. 
The yellow-shaded areas correspond to the energy gap.
}
\label{fig:4} 
\end{figure*}

Figures~\ref{fig:4}(a) and \ref{fig:4}(d) show the band structures for $h = 5$ and $\alpha = 1$.
Here, a relatively large value of $h$ is introduced so as to open a gap and realize an insulating state at half filling.
Figures~\ref{fig:4}(b) and \ref{fig:4}(e) show the chemical potential dependence of $\chi^{T_0}$ at $T = 0.1$ and $0.01$, respectively.
The magnitude of $\chi^{T_0}$ remains of the same order in both metallic and insulating regimes.
Figures~\ref{fig:4}(c) and \ref{fig:4}(f) present the corresponding behavior of $\tilde{T}_0$ at $T = 0.1$ and $0.01$.
Similarly, $\tilde{T}_0$ exhibits a comparable magnitude across the two regimes.

On the metallic side of Fig.~\ref{fig:4}, pronounced structures in $\chi^{T_0}$ appear at chemical potentials associated with band features near the high-symmetry points X, M, $\Gamma$, R, and A.
These structures are likely enhanced by small interband energy separations, which enter the denominators in Eqs.~(\ref{eq:def1})--(\ref{eq:def3}).

In the insulating regime, $\chi^{T_0}$ is independent of $\mu$, whereas $\tilde{T}_0$ exhibits a linear dependence on $\mu$.
This difference originates from the distinct $\mu$ dependence of the Fermi distribution function $f_n$ and the grand potential density $\mathcal{G}_n$.
At zero temperature, $\sum_n f_n$ reduces to a sum over occupied states, $\sum_n^{\text{occ}}$, while $\sum_n \mathcal{G}_n$ becomes $\sum_n^{\text{occ}} (\epsilon_n - \mu)$. 
Consequently, $\chi^{T_0}$ and $\tilde{T}_0$ exhibit distinct dependencies on the chemical potential.

This linear dependence does not indicate an ambiguity in the definition of $\tilde{T}_0$, since its additive constant is fixed by the empty-system reference at $\mu=-\infty$.
Rather, $\tilde{T}_0$ should be regarded as a bulk quantity in the grand-canonical formulation with the chemical potential specified, while $\chi^{T_0}$ is the corresponding linear-response coefficient.
This relation is analogous to Streda-type relations, in which the chemical-potential derivative of a bulk thermodynamic quantity is related to a measurable response.

\section{Summary}\label{Summary}
We have developed a theoretical framework for characterizing the MTM in periodic crystals from the viewpoint of response theory. 
Motivated by the fact that the MTM is a rank-0 time-reversal-odd scalar quantity, we focused on the completely antisymmetric component of a rank-3 response tensor and formulated a gauge-invariant scheme to extract the MTM from the response of relativistic electric polarization to a magnetic field gradient. 
By employing the Kubo formalism, we derived the response tensor associated with the magnetic-field-gradient correction to relativistic electric polarization. 
The obtained expression was shown to be described by geometric quantities defined in an extended parameter space spanned by crystal momentum, magnetic field, and electric field. 
This indicates that the MTM-related response is intrinsically linked to the quantum geometry of electronic states.
Since the resulting expression is gauge invariant and expressed in terms of the Bloch wave functions, it can be used in first-principles calculations to evaluate the completely antisymmetric component of the electric polarization induced by a magnetic field gradient.
Note that, in the absence of relativistic contributions, this component vanishes according to \Eq{eq: thermo_relation}.
We also established a bulk formulation of the MTM by integrating the response function with respect to the chemical potential. 
The resulting expression for the MTM is written as a linear combination of couplings among relativistic polarization, velocity, spin, and dipolar quantities such as the magnetic toroidal, electric, and electric toroidal dipoles. 
Furthermore, our results for characterizing MTM in a crystal are applicable to both metals and insulators at any temperatures.
Finally, we demonstrate the formulation in a minimal antiferromagnetic model, where the MTM becomes finite. 

The present results provide a practical route to characterize the MTMs in periodic crystals and clarify their microscopic origin in terms of extended quantum geometry. 
They also offer a basis for future investigations of MTM-related responses in realistic materials.
As candidate materials exhibiting these responses, not only Co$_2$SiO$_4$\cite{hayashida2025electric,kato2026electric} but also KMnF$_3$\cite{KNIGHT2020155935}, MnV$_2$O$_4$\cite{garlea2008magnetic}, Er$_2$Cu$_2$O$_5$\cite{garca1991complex}, Ho$_2$Ge$_2$O$_7$\cite{morosan2008structure}, and Mn$_2$FeMoO$_6$\cite{li2014magnetic} are promising candidates.

\begin{acknowledgments}
This research was supported by JSPS KAKENHI Grants Numbers JP22H00101, JP22H01183, JP23H04869, JP23K03288, and by JST CREST (JPMJCR23O4) and JST FOREST (JPMJFR2366). 
\end{acknowledgments}

\appendix
  \begin{widetext}
\section{Derivation of \Eq{eq:responce}}\label{derivation}
We derive \Eq{eq:responce} expanding the correlation function~\Eq{eq:correlation}
\begin{align}
\chi(\vb*{q}) = 
\lim_{\delta \rightarrow 0} \chi_{\tilde{P}_k, M_i}(\vb*{q}, 0) 
= -g\mu_\text{B} \xi\sum_{n,m}^{} &\int \frac{d^dk}{(2\pi)^d} \bra{n\vb*{k}_-} 
\frac{1}{2}\left(\hat{\tilde{r}}_k (\vb*{k}_+) + \hat{\tilde{r}}_k (\vb*{k}_-) \right) 
\ket{m\vb*{k}_+}\bra{m\vb*{k}_+} \hat{s}_i \ket{n\vb*{k}_-} \frac{f_{n\vb*{k}_-} - f_{m\vb*{k}_+}}{\epsilon_{n\vb*{k}_-} - \epsilon_{m\vb*{k}_+} },
\end{align}
We consider an expansion up to first order in the wave vector $q$. 
For $n \neq m$, this can be expanded as follows: 
\begin{align}
  \bra{n\vb*{k}_-} \frac{1}{2}\left(\hat{\tilde{r}}_k (\vb*{k}_+) + \hat{\tilde{r}}_k (\vb*{k}_-) \right) \ket{m\vb*{k}_+} 
  &\approx \tilde{r}^k_{nm} + \frac{q_j}{2}\left( -\bra{\partial_j n} \tilde{\hat{r}}_k \ket{m} + \bra{n} \tilde{\hat{r}}_k \ket{\partial_j m} \right), \nonumber\\
  \bra{m\vb*{k}+} \hat{s}_i \ket{n\vb*{k}+} &\approx s^i_{nm} + \frac{q_j}{2} \left( \bra{\partial_j m} \hat{s}_i \ket{n} - \bra{m} \hat{s}_i \ket{\partial_j n} \right), \nonumber\\
  \frac{f_{n\vb*{k}_-} - f_{m\vb*{k}_+}}{\epsilon_{n\vb*{k}_-} - \epsilon_{m\vb*{k}_+} } & \approx \frac{f_{nm}}{\epsilon_{nm}} - \frac{q_j}{2} \frac{1}{\epsilon_{nm}} \left(\partial_j \tilde{f}_{nm} - \frac{\partial_j \tilde{\epsilon}_{nm}}{\epsilon_{nm}}f_{nm} \right).
\end{align}
Here, we use the abbreviations $\tilde{\epsilon}_{nm} = \epsilon_n + \epsilon_m$, $\tilde{f}_{nm} = f_{n\vb*{k}} + f_{m\vb*{k}}$, $f_{nm} = f_{n\vb*{k}} - f_{m\vb*{k}}$.
From these equations, the first-order contribution of the $n \neq m$ case as
\begin{align}
  \chi^{n \neq m}(\vb*{q})&=  -g\mu_\text{B} \xi \sum_{n}^{} \sum_{m}^{n \neq m} \int \frac{d^dk}{(2\pi)^d}\frac{q_j}{2}
  \Bigg\{
    \tilde{r}^k_{nm} \left( \bra{\partial_j m} \hat{s}_i \ket{n} - \bra{m} \hat{s}_i \ket{\partial_j n} \right) \frac{f_{nm}}{\epsilon_{nm}} + \left( -\bra{\partial_j n} \tilde{\hat{r}}_k \ket{m} + \bra{n} \tilde{\hat{r}}_k \ket{\partial_j m} \right)s^i_{mn} \frac{f_{nm}}{\epsilon_{nm}} \nonumber\\
    &- \tilde{r}^k_{nm} s^i_{mn}\frac{1}{\epsilon_{nm}} \left(\partial_j \tilde{f}_{nm} - \frac{\partial_j \tilde{\epsilon}_{nm}}{\epsilon_{nm}}f_{nm} \right) 
  \Bigg\} .
\end{align}
The first and second terms inside the curly brackets can be transformed as follows
\begin{align}
  \sum_{n}^{} \sum_{m}^{n \neq m} 
    &\left\{\tilde{r}^k_{nm} \left( \bra{\partial_j m} \hat{s}_i \ket{n} - \bra{m} \hat{s}_i \ket{\partial_j n} \right) + \left( -\bra{\partial_j n} \tilde{\hat{r}}_k \ket{m} + \bra{n} \tilde{\hat{r}}_k \ket{\partial_j m} \right)s^i_{mn}\right\} \frac{f_{nm}}{\epsilon_{nm}} \nonumber\\
    &= 2i \sum_{n}^{} \sum_{m}^{n \neq m}
    \Bigg\{
    -\Im\left[ \frac{\tilde{r}^k_{nm} v^j_{mn}}{\epsilon^2_{nm}} \right](s^i_n + s^i_m) 
    -\Im\left[ \frac{v^j_{nm} s^i_{mn}}{\epsilon^2_{nm}} \right](\tilde{r}^k_n + \tilde{r}^k_m) \nonumber\\
    &+\sum_{l}^{l \neq n, m} \Im\Bigg[ \frac{s^i_{ln}v^j_{ml}\tilde{r}^k_{nm} - s^i_{nm}v^j_{ml}\tilde{r}^k_{ln}}{\epsilon_{nm}\epsilon_{ml}} + \frac{s^i_{ml}v^j_{ln}\tilde{r}^k_{nm} - s^i_{nm}v^j_{ln}\tilde{r}^k_{ml}}{\epsilon_{nm}\epsilon_{ln}} \Bigg]
    \Bigg\} f_n .
\end{align}
Here, we use $\bra{n}\ket{\partial_j m} = -v^j_{nm} / \epsilon_{nm}$.
 The third term is 
 \begin{align}
  - \sum_{n}^{} \sum_{m}^{n \neq m} \tilde{r}^k_{nm} s^i_{mn}\frac{1}{\epsilon_{nm}} \left(\partial_j \tilde{f}_{nm} - \frac{\partial_j \tilde{\epsilon}_{nm}}{\epsilon_{nm}}f_{nm} \right)
  &= 2i\sum_{n}^{} \sum_{m}^{n \neq m} 
  \Bigg\{
   \Im\left[ \frac{\tilde{r}^k_{nm} s^i_{mn}}{\epsilon_{nm}} \right]v^j_n f'_n
  - \Im\left[ \frac{\tilde{r}^k_{nm} s^i_{mn}}{\epsilon^2_{nm}} \right](v^j_n + v^j_m)f_n  
  \Bigg\} .
 \end{align}
 From these equation, 
 \begin{align}\label{eq:intra}
  \chi^{n \neq m}(\vb*{q})
  = -g\mu_\text{B} \xi  \int \frac{d^dk}{(2\pi)^d}iq_j \sum_{n}^{} &\sum_{m}^{n \neq m}
  \Bigg[
  -\frac{1}{2}\Bigg\{(s^i_n + s^i_m) \Omega^{v_j, \tilde{r}_k}_{nm} + (v^j_n + v^j_m) \Omega^{\tilde{r}_k, s_i}_{nm} + (\tilde{r}^k_n + \tilde{r}^k_m) \Omega^{s_i, v_j}_{nm} \Bigg\}f_n \nonumber\\
&- v^j_n m^{\tilde{r}_k, s_i}_{nm}f'_n + \sum_{l}^{\neq n, m}X^{ijk}_{nml}f_n
\Bigg] .
 \end{align}
 In the case of $n = m$, 
 \begin{align}
  \bra{n\vb*{k}-} 
  \frac{1}{2}\left(\hat{\tilde{r}}_k (\vb*{k}_+) + \hat{\tilde{r}}_k (\vb*{k}_-) \right)\ket{m\vb*{k}+} &\approx \tilde{r}^k_{n} + \frac{q_j}{2}\left( -\bra{\partial_j n} \tilde{\hat{r}}_k \ket{n} + \bra{n} \tilde{\hat{r}}_k \ket{\partial_j n} \right), \nonumber\\
  \bra{m\vb*{k}+} \hat{s}_i \ket{n\vb*{k}+} &\approx s^i_{n} + \frac{q_j}{2} \left( \bra{\partial_j n} \hat{s}_i \ket{n} - \bra{n} \hat{s}_i \ket{\partial_j n} \right), \nonumber\\
  \frac{f_{n\vb*{k}_-} - f_{m\vb*{k}_+}}{\epsilon_{n\vb*{k}_-} - \epsilon_{m\vb*{k}_+} } & \approx f'_n .
 \end{align}
 From these equations, the first-order contribution of the $n = m$ case as
 \begin{align}\label{eq:inter}
    \chi^{n = m}(\vb*{q}) 
  &=  -g\mu_\text{B} \xi \int \frac{d^dk}{(2\pi)^d}\frac{q_j}{2}\sum_{n}^{}
  \Bigg\{
     \tilde{r}^k_{n}\left( \bra{\partial_j n} \hat{s}_i \ket{n} - \bra{n} \hat{s}_i \ket{\partial_j n} \right) + \left( -\bra{\partial_j n} \tilde{\hat{r}}_k \ket{n} + \bra{n} \tilde{\hat{r}}_k \ket{\partial_j n} \right)s^i_{n}
  \Bigg\} f'_n \nonumber\\
  &= -g\mu_\text{B} \xi \int \frac{d^dk}{(2\pi)^d}iq_j \sum_{n}^{} \sum_{m}^{n \neq m}
  \left\{\Im\left[ \frac{\tilde{r}^k_{nm} v^j_{mn}}{\epsilon_{nm}} \right]s^i_n f'_n + \Im\left[ \frac{v^j_{nm} s^i_{mn}}{\epsilon_{nm}} \right]\tilde{r}^k_n f'_n\right\} \nonumber\\
  &= -g\mu_\text{B} \xi  \int \frac{d^dk}{(2\pi)^d}iq_j \sum_{n}^{} \sum_{m}^{n \neq m} \left(-s^i_n m^{v_j, \tilde{r}_k}_{nm}  -\tilde{r}^k_n m^{s_i, v_j}_{nm}\right)f'_n .
 \end{align}
 Combining \Eq{eq:intra} and \Eq{eq:inter}, we obtain the $\chi_{ijk}$ as 
 \begin{align}
  \chi_{ijk} 
  &= \lim_{\vb*{q} \rightarrow 0} -i\partial_{q_j} \chi(\vb*{q}) \nonumber\\
  &= -g\mu_{\text{B}}\xi\int_{}^{}\frac{d^dk}{(2\pi)^d} \sum_{n}^{} 
  \Bigg[ 
\sum_{m}^{\neq n}-\frac{1}{2}\Bigg\{(s^i_n + s^i_m) \Omega^{v_j, \tilde{r}_k}_{nm} + (v^j_n + v^j_m) \Omega^{\tilde{r}_k, s_i}_{nm} + (\tilde{r}^k_n + \tilde{r}^k_m) \Omega^{s_i, v_j}_{nm} \Bigg\}f_n \nonumber\\
&- \Bigg\{s^i_n m^{v_j, \tilde{r}_k}_{nm} + v^j_n m^{\tilde{r}_k, s_i}_{nm}+ \tilde{r}^k_n m^{s_i, v_j}_{nm} \Bigg\}f'_n +\sum_{m}^{\neq n} \sum_{l}^{\neq n, m}X^{ijk}_{nml}f_n
  \Bigg],
 \end{align}
which corresponds to \Eq{eq:responce}.
\end{widetext}

\section{Expression near degenerate points}\label{degeneracy}
To perform the numerical calculations, we begin with an equivalent form of \Eq{eq:responce}: 
\begin{widetext}
   \begin{align}\label{eq:deg}
  \chi_{ijk} = -g\mu_{\text{B}}\xi\int_{}^{}\frac{d^dk}{(2\pi)^d} \sum_{n}^{}\sum_{m}^{\neq n}
  \Bigg[ \left\{s^i_n\Im\left[ v^j_{nm}{\tilde{r}}^k_{mn} \right] + v^j_n \Im\left[{\tilde{r}}^k_{nm} {s}^i_{mn} \right] + \tilde{r}^k_n\Im\left[ s^i_{nm} v^j_{mn} \right]\right\}P_{nm} 
  +\sum_{l}^{\neq n, m}\Im\left[ s^i_{nm} v^j_{ml}{\tilde{r}}^k_{ln}\right]Q_{nml} 
  \Bigg],
\end{align}
\end{widetext}
where
\begin{align}
  P_{nm} &= \frac{f_{nm}}{\epsilon^2_{nm}}-\frac{f'_n}{\epsilon_{nm}}, \nonumber\\
  Q_{nml} &= \frac{f_{nm}}{\epsilon_{nm}\epsilon_{ml}} - \frac{f_{ln}}{\epsilon_{ml}\epsilon_{ln}}.
\end{align}
$P_{nm}$ has the asymptotic form near the degeneracy point ($\epsilon_{n} = \epsilon_{m}$)
\begin{align}
 P_{nm} =  - \frac{1}{2}f''_n + \frac{1}{6}f'''_n\epsilon_{nm} + O(\epsilon^2_{nm}) .
\end{align}
For $Q_{nml}$, cases should be classified according to the type of degeneracy:
\begin{itemize}
    \item[(i)]   $\epsilon_{n} = \epsilon_{m}$ and $\epsilon_{m} \neq \epsilon_{l}$ ,
    \item[(ii)]  $\epsilon_{n} \neq \epsilon_{m}$ and $\epsilon_{m} = \epsilon_{l}$ ,
    \item[(iii)] $\epsilon_{n} = \epsilon_{m} = \epsilon_{l}$ .
\end{itemize}
In the case of (i),
\begin{align}
 Q_{nml} = \left(\frac{f_{ln}}{\epsilon^2_{ln}} - \frac{f'_n}{\epsilon_{ln}}\right) + \left( \frac{f''_n}{2\epsilon_{ln}} + \frac{f'_n}{\epsilon^2_{ln}} - \frac{f_{ln}}{\epsilon^3_{ln}}\right)\epsilon_{nm} + O(\epsilon^2_{nm}) .
\end{align}
In the case of (ii),
\begin{align}
  Q_{nml} = \left(\frac{f_{nm}}{\epsilon^2_{nm}} - \frac{f'_m}{\epsilon_{nm}}\right) + \left(\frac{f''_m}{2\epsilon_{nm}} + \frac{f'_m}{\epsilon^2_{nm}} - \frac{f_{nm}}{\epsilon^3_{nm}}\right)\epsilon_{ml} + O(\epsilon^2_{ml}) .
\end{align}
In the case of (iii),
\begin{align}
Q_{nml} = \frac{1}{2}f''_m - \frac{1}{6}f'''_m\epsilon_{nm} + O(\epsilon^2_{nm}) .
\end{align}
Thus, Eq.~(\ref{eq:deg}) remains well defined even at degenerate points.

\bibliography{main}

@article{yamanaka2025nonlinear,
  title={{Nonlinear Nonreciprocal Electric Conductivity Driven by Magnetic Field Gradients}},
  author={Yamanaka, Taisei and Ihara, Yoshihiko and Hayami, Satoru},
  journal={J. Phys. Soc. Jpn.},
  volume={94},
  number={8},
  pages={083705},
  year={2025},
  publisher={The Physical Society of Japan},
  doi={10.7566/JPSJ.94.083705}
}

@article{RevModPhys.82.1959,
  title = {{Berry phase effects on electronic properties}},
  author = {Xiao, Di and Chang, Ming-Che and Niu, Qian},
  journal = {Rev. Mod. Phys.},
  volume = {82},
  issue = {3},
  pages = {1959--2007},
  numpages = {0},
  year = {2010},
  month = {Jul},
  publisher = {American Physical Society},
  doi = {10.1103/RevModPhys.82.1959},
  url = {https://link.aps.org/doi/10.1103/RevModPhys.82.1959}
}

@article{PhysRevB.53.7010,
  title = {{Berry phase, hyperorbits, and the Hofstadter spectrum: Semiclassical dynamics in magnetic Bloch bands}},
  author = {Chang, Ming-Che and Niu, Qian},
  journal = {Phys. Rev. B},
  volume = {53},
  issue = {11},
  pages = {7010--7023},
  numpages = {0},
  year = {1996},
  month = {Mar},
  publisher = {American Physical Society},
  doi = {10.1103/PhysRevB.53.7010},
  url = {https://link.aps.org/doi/10.1103/PhysRevB.53.7010}
}

@article{sato2026conductivity,
author = {Sato, Takumi and Hayami, Satoru},
title = {Thermodynamic Multipoles and Dissipative Conductivities in Metallic Systems},
journal = {Ann. Phys. (Berlin)},
volume = {538},
number = {8},
pages = {e70251},
doi = {https://doi.org/10.1002/andp.70251},
url = {https://onlinelibrary.wiley.com/doi/abs/10.1002/andp.70251},
eprint = {https://onlinelibrary.wiley.com/doi/pdf/10.1002/andp.70251},
year = {2026}
}

@article{Hayami_doi:10.7566/JPSJ.91.113702,
author = {Hayami ,Satoru and Oiwa ,Rikuto and Kusunose ,Hiroaki},
title={{Electric Ferro-Axial Moment as Nanometric Rotator and Source of Longitudinal Spin Current}},
journal = {J. Phys. Soc. Jpn.},
volume = {91},
number = {11},
pages = {113702},
year = {2022},
doi = {10.7566/JPSJ.91.113702},
URL = {https://doi.org/10.7566/JPSJ.91.113702},
}

@article{shitade2025intrinsic,
  title = {{Intrinsic spin accumulation in the magnetic spin Hall effect}},
  author = {Shitade, Atsuo},
  journal = {Phys. Rev. B},
  volume = {112},
  issue = {17},
  pages = {174431},
  numpages = {14},
  year = {2025},
  month = {Nov},
  publisher = {American Physical Society},
  doi = {10.1103/gxpm-2gkq},
  url = {https://link.aps.org/doi/10.1103/gxpm-2gkq}
}

@article{schmid2008some,
  author={Schmid, Hans},
  journal={J. Phys.: Condens. Matter},
  volume={20},
  number={43},
  pages={434201},
  year={2008},
  publisher={IOP Publishing},
  doi={10.1088/0953-8984/20/43/434201}
}

@article{schmid2001ferrotoroidics,
  title={{On ferrotoroidics and electrotoroidic, magnetotoroidic and piezotoroidic effects}},
  author={Schmid, Hans},
  journal={Ferroelectrics},
  volume={252},
  number={1},
  pages={41--50},
  year={2001},
  publisher={Taylor \& Francis},
  doi={https://doi.org/10.1080/00150190108016239}
}

@article{kuniyoshi2026theory,
author = {Kuniyoshi ,Shingo and Oiwa ,Rikuto and Hayami ,Satoru},
title = {Theory of Many-Body Multipole Operators in Single-Centered Electron Systems: Two-Body Toroidal Monopoles in Spinless Orbitals},
journal = {J. Phys. Soc. Jpn.},
volume = {95},
number = {7},
pages = {073704},
year = {2026},
doi = {10.7566/JPSJ.95.073704},
URL = {https://doi.org/10.7566/JPSJ.95.073704
}
}

@article{curie1894symetrie,
  title={{Sur la sym{\'e}trie dans les ph{\'e}nom{\`e}nes physiques, sym{\'e}trie d'un champ {\'e}lectrique et d'un champ magn{\'e}tique}},
  author={Curie, Pierre},
  journal={J. Phys. Theor. Appl.},
  volume={3},
  number={1},
  pages={393--415},
  year={1894}
}

@article{Spaldin2005renaissance,
  title={{The renaissance of magnetoelectric multiferroics}},
  author={Spaldin, Nicola A and Fiebig, Manfred},
  journal={Science},
  volume={309},
  number={5733},
  pages={391--392},
  year={2005},
  publisher={American Association for the Advancement of Science},
  doi = {10.1126/science.1113357},
}

@article{Fiebig0022-3727-38-8-R01,
  author={Manfred Fiebig},
  title={{Revival of the magnetoelectric effect}},
  journal={J. Phys. D: Appl. Phys.},
  volume={38},
  number={8},
  pages={R123},
  year={2005},
  doi = {10.1088/0022-3727/38/8/R01},
  }

@article{Chen_PhysRevLett.112.017205,
  title = {{Anomalous {Hall} Effect Arising from Noncollinear Antiferromagnetism}},
  author = {Chen, Hua and Niu, Qian and MacDonald, A. H.},
  journal = {Phys. Rev. Lett.},
  volume = {112},
  issue = {1},
  pages = {017205},
  numpages = {5},
  year = {2014},
  month = {Jan},
  publisher = {American Physical Society},
  doi = {10.1103/PhysRevLett.112.017205},
  url = {https://link.aps.org/doi/10.1103/PhysRevLett.112.017205}
}

@article{Hayami_PhysRevB.90.081115,
  title = {Spontaneous parity breaking in spin-orbital coupled systems},
  author = {Hayami, Satoru and Kusunose, Hiroaki and Motome, Yukitoshi},
  journal = {Phys. Rev. B},
  volume = {90},
  issue = {8},
  pages = {081115},
  numpages = {5},
  year = {2014},
  month = {Aug},
  publisher = {American Physical Society},
  doi = {10.1103/PhysRevB.90.081115},
}

@article{vsmejkal2020crystal,
  title={{Crystal time-reversal symmetry breaking and spontaneous {Hall} effect in collinear antiferromagnets}},
  author={{\v{S}}mejkal, Libor and Gonz{\'a}lez-Hern{\'a}ndez, Rafael and Jungwirth, T and Sinova, J},
  journal={Sci. Adv.},
  volume={6},
  number={23},
  pages={eaaz8809},
  year={2020},
  publisher={American Association for the Advancement of Science},
  doi={10.1126/sciadv.aaz8809},
}

@article{Chen_PhysRevB.106.024421,
  title = {{Electronic chiralization as an indicator of the anomalous {Hall} effect in unconventional magnetic systems}},
  author = {Chen, Hua},
  journal = {Phys. Rev. B},
  volume = {106},
  issue = {2},
  pages = {024421},
  numpages = {13},
  year = {2022},
  month = {Jul},
  publisher = {American Physical Society},
  doi = {10.1103/PhysRevB.106.024421},
  url = {https://link.aps.org/doi/10.1103/PhysRevB.106.024421}
}

@article{Hayami_PhysRevB.103.L180407,
  title = {{Essential role of the anisotropic magnetic dipole in the anomalous {Hall} effect}},
  author = {Hayami, Satoru and Kusunose, Hiroaki},
  journal = {Phys. Rev. B},
  volume = {103},
  issue = {18},
  pages = {L180407},
  numpages = {5},
  year = {2021},
  month = {May},
  publisher = {American Physical Society},
  doi = {10.1103/PhysRevB.103.L180407},
  url = {https://link.aps.org/doi/10.1103/PhysRevB.103.L180407}
}

@article{Sivadas_PhysRevLett.117.267203,
  title = {{Gate-Controllable Magneto-optic Kerr Effect in Layered Collinear Antiferromagnets}},
  author = {Sivadas, Nikhil and Okamoto, Satoshi and Xiao, Di},
  journal = {Phys. Rev. Lett.},
  volume = {117},
  issue = {26},
  pages = {267203},
  numpages = {5},
  year = {2016},
  month = {Dec},
  publisher = {American Physical Society},
  doi = {10.1103/PhysRevLett.117.267203},
  url = {https://link.aps.org/doi/10.1103/PhysRevLett.117.267203}
}

@article{Solovyev_PhysRevB.55.8060,
  title = {{Magneto-optical effect in the weak ferromagnets ${\mathrm{LaMO}}_{3}$ (M$=$ Cr, Mn, and Fe)}},
  author = {Solovyev, I. V.},
  journal = {Phys. Rev. B},
  volume = {55},
  issue = {13},
  pages = {8060--8063},
  numpages = {0},
  year = {1997},
  month = {Apr},
  publisher = {American Physical Society},
  doi = {10.1103/PhysRevB.55.8060},
  url = {https://link.aps.org/doi/10.1103/PhysRevB.55.8060}
}

@article{Loss_PhysRevB.45.13544,
  title = {{Persistent currents from {Berry's} phase in mesoscopic systems}},
  author = {Loss, Daniel and Goldbart, Paul M.},
  journal = {Phys. Rev. B},
  volume = {45},
  issue = {23},
  pages = {13544--13561},
  numpages = {0},
  year = {1992},
  month = {Jun},
  publisher = {American Physical Society},
  doi = {10.1103/PhysRevB.45.13544},
}

@article{suzuki2018first,
	author = {Suzuki, Michi-To and Ikeda, Hiroaki and Oppeneer, Peter M},
	doi = {10.7566/JPSJ.87.041008},
	journal = {J. Phys. Soc. Jpn.},
	number = {4},
	pages = {041008},
	publisher = {The Physical Society of Japan},
	title={{First-principles theory of magnetic multipoles in condensed matter systems}},
	volume = {87},
	year = {2018}}

@article{Kusunose_PhysRevB.107.195118,
	author = {Kusunose, Hiroaki and Oiwa, Rikuto and Hayami, Satoru},
	doi = {10.1103/PhysRevB.107.195118},
	issue = {19},
	journal = {Phys. Rev. B},
	month = {May},
	numpages = {14},
	pages = {195118},
	publisher = {American Physical Society},
	title={{Symmetry-adapted modeling for molecules and crystals}},
	url = {https://link.aps.org/doi/10.1103/PhysRevB.107.195118},
	volume = {107},
	year = {2023}}

@article{kusunose2022generalization,
	author = {Hiroaki Kusunose and Satoru Hayami},
	doi = {10.1088/1361-648x/ac9209},
	journal = {J. Phys.: Condens. Matter},
	month = {sep},
	number = {46},
	pages = {464002},
	publisher = {{IOP} Publishing},
	title={{Generalization of microscopic multipoles and cross-correlated phenomena by their orderings}},
	url = {https://doi.org/10.1088/1361-648x/ac9209},
	volume = {34},
	year = 2022}

@article{Spaldin_0953-8984-20-43-434203,
	author = {Nicola A Spaldin and Manfred Fiebig and Maxim Mostovoy},
	doi = {10.1088/0953-8984/20/43/434203},
	journal = {J. Phys.: Condens. Matter},
	number = {43},
	pages = {434203},
	title={{The toroidal moment in condensed-matter physics and its relation to the magnetoelectric effect}},
	volume = {20},
	year = {2008}}

@article{tokura2018nonreciprocal,
	author = {Tokura, Yoshinori and Nagaosa, Naoto},
	doi = {10.1038/s41467-018-05759-4},
	journal = {Nat. Commun.},
	number = {1},
	pages = {3740},
	publisher = {Nature Publishing Group},
	title={{Nonreciprocal responses from non-centrosymmetric quantum materials}},
	volume = {9},
	year = {2018}}

@article{Suzuki_PhysRevB.105.075201,
	author = {Suzuki, Yuta},
	doi = {10.1103/PhysRevB.105.075201},
	issue = {7},
	journal = {Phys. Rev. B},
	month = {Feb},
	numpages = {11},
	pages = {075201},
	publisher = {American Physical Society},
	title={{Tunneling spin current in systems with spin degeneracy}},
	url = {https://link.aps.org/doi/10.1103/PhysRevB.105.075201},
	volume = {105},
	year = {2022}}

@article{popov1999magnetic,
	author = {Popov, Yu F and Kadomtseva, AM and Belov, DV and Vorob'ev, GP and Zvezdin, AK},
	doi = {10.1134/1.568032},
	journal = {J. Exp. Theor. Phys. Lett.},
	number = {4},
	pages = {330--335},
	publisher = {Springer},
	title={{Magnetic-field-induced toroidal moment in the magnetoelectric Cr2O3}},
	volume = {69},
	year = {1999}}

@article{EdererPhysRevB.76.214404,
	author = {Ederer, Claude and Spaldin, Nicola A.},
	doi = {10.1103/PhysRevB.76.214404},
	issue = {21},
	journal = {Phys. Rev. B},
	month = {Dec},
	numpages = {13},
	pages = {214404},
	publisher = {American Physical Society},
	title={{Towards a microscopic theory of toroidal moments in bulk periodic crystals}},
	volume = {76},
	year = {2007}}

@article{thole2018magnetoelectric,
	author = {Th{\"o}le, Florian and Spaldin, Nicola A},
	doi = {10.1098/rsta.2017.0450},
	journal = {Philos. Trans. R. Soc. A},
	number = {2134},
	pages = {20170450},
	publisher = {The Royal Society Publishing},
	title={{Magnetoelectric multipoles in metals}},
	volume = {376},
	year = {2018}}

@article{Hayami_PhysRevB.106.014420,
	author = {Hayami, Satoru and Yatsushiro, Megumi},
	doi = {10.1103/PhysRevB.106.014420},
	issue = {1},
	journal = {Phys. Rev. B},
	month = {Jul},
	numpages = {6},
	pages = {014420},
	publisher = {American Physical Society},
	title={{Nonlinear nonreciprocal transport in antiferromagnets free from spin-orbit coupling}},
	url = {https://link.aps.org/doi/10.1103/PhysRevB.106.014420},
	volume = {106},
	year = {2022}}

@article{yatsushiro2022analysis,
	author = {Yatsushiro, Megumi and Oiwa, Rikuto and Kusunose, Hiroaki and Hayami, Satoru},
	journal = {Phys. Rev. B},
	number = {15},
	pages = {155157},
	publisher = {APS},
	title={{Analysis of model-parameter dependences on the second-order nonlinear conductivity in PT-symmetric collinear antiferromagnetic metals with magnetic toroidal moment on zigzag chains}},
	volume = {105},
	year = {2022},
        doi = {doi.org/10.1103/PhysRevB.105.155157},
        url={https://doi.org/10.1103/PhysRevB.105.155157}
}

@article{nagaosa2024nonreciprocal,
	author = {Nagaosa, Naoto and Yanase, Youichi},
	journal = {Annu. Rev. Condens. Matter Physis},
	number = {1},
	pages = {63--83},
	publisher = {Annual Reviews},
	title={{Nonreciprocal transport and optical phenomena in quantum materials}},
	volume = {15},
	year = {2024},
        doi = {doi.org/10.1146/annurev-conmatphys-032822-033734},
        url = {https://doi.org/10.1146/annurev-conmatphys-032822-033734}
}

@article{hayami2024unified,
	author = {Hayami, Satoru and Kusunose, Hiroaki},
	journal = {J. Phys. Soc. Jpn.},
	number = {7},
	pages = {072001},
	publisher = {The Physical Society of Japan},
	title={{Unified description of electronic orderings and cross correlations by complete multipole representation}},
	volume = {93},
	year = {2024}}

@article{hayami2022nonlinear,
  title = {{Nonlinear spin Hall effect in $\mathcal{PT}$-symmetric collinear magnets}},
  author = {Hayami, Satoru and Yatsushiro, Megumi and Kusunose, Hiroaki},
  journal = {Phys. Rev. B},
  volume = {106},
  issue = {2},
  pages = {024405},
  numpages = {9},
  year = {2022},
  month = {Jul},
  publisher = {American Physical Society},
  doi = {10.1103/PhysRevB.106.024405},
  url = {https://link.aps.org/doi/10.1103/PhysRevB.106.024405}
}

@article{watanabe2020nonlinear,
	author = {Watanabe, Hikaru and Yanase, Youichi},
	journal = {Phys. Rev. Res.},
	number = {4},
	pages = {043081},
	publisher = {APS},
	title={{Nonlinear electric transport in odd-parity magnetic multipole systems: Application to Mn-based compounds}},
	volume = {2},
	year = {2020},
        doi = {10.1103/PhysRevResearch.2.043081},
        url = {https://link.aps.org/doi/10.1103/PhysRevResearch.2.043081}
}

@article{wakatsuki2017nonreciprocal,
	author = {Wakatsuki, Ryohei and Saito, Yu and Hoshino, Shintaro and Itahashi, Yuki M and Ideue, Toshiya and Ezawa, Motohiko and Iwasa, Yoshihiro and Nagaosa, Naoto},
	journal = {Science advances},
	number = {4},
	pages = {e1602390},
	publisher = {American Association for the Advancement of Science},
	title={{Nonreciprocal charge transport in noncentrosymmetric superconductors}},
	volume = {3},
	year = {2017},
        doi = {10.1126/sciadv.1602390}
}

@article{rikken1997observation,
	author = {Rikken, GLJA and Raupach, E},
	journal = {Nature},
	number = {6659},
	pages = {493--494},
	publisher = {Nature Publishing Group UK London},
	title={{Observation of magneto-chiral dichroism}},
	volume = {390},
	year = {1997},
        doi = {doi.org/10.1038/37323},
        url = {https://doi.org/10.1038/37323}
}

@article{shitade2019theory,
  title={{Theory of spin magnetic quadrupole moment and temperature-gradient-induced magnetization}},
  author = {Shitade, Atsuo and Daido, Akito and Yanase, Youichi},
  journal = {Phys. Rev. B},
  volume = {99},
  issue = {2},
  pages = {024404},
  numpages = {12},
  year = {2019},
  month = {Jan},
  publisher = {American Physical Society},
  doi = {10.1103/PhysRevB.99.024404},
  url = {https://link.aps.org/doi/10.1103/PhysRevB.99.024404}
}

@article{shitade2018theory,
  title={{Theory of orbital magnetic quadrupole moment and magnetoelectric susceptibility}},
  author = {Shitade, Atsuo and Watanabe, Hikaru and Yanase, Youichi},
  journal = {Phys. Rev. B},
  volume = {98},
  issue = {2},
  pages = {020407},
  numpages = {6},
  year = {2018},
  month = {Jul},
  publisher = {American Physical Society},
  doi = {10.1103/PhysRevB.98.020407},
  url = {https://link.aps.org/doi/10.1103/PhysRevB.98.020407}
}

@article{daido2020thermodynamic,
  title={{Thermodynamic approach to electric quadrupole moments}},
  author = {Daido, Akito and Shitade, Atsuo and Yanase, Youichi},
  journal = {Phys. Rev. B},
  volume = {102},
  issue = {23},
  pages = {235149},
  numpages = {12},
  year = {2020},
  month = {Dec},
  publisher = {American Physical Society},
  doi = {10.1103/PhysRevB.102.235149},
  url = {https://link.aps.org/doi/10.1103/PhysRevB.102.235149}
}

@article{oike2025thermodynamic,
  title={{Thermodynamic formulation of the spin magnetic octupole moment in bulk crystals}},
  author = {\ifmmode \bar{O}\else \={O}\fi{}ik\'e, Jun and Peters, Robert and Shinada, Koki},
  journal = {Phys. Rev. B},
  volume = {112},
  issue = {13},
  pages = {134412},
  numpages = {15},
  year = {2025},
  month = {Oct},
  publisher = {American Physical Society},
  doi = {10.1103/frq1-9xx7},
  url = {https://link.aps.org/doi/10.1103/frq1-9xx7}
}

@article{gao2018microscopic,
  title={{Microscopic theory of spin toroidization in periodic crystals}},
  author = {Gao, Yang and Vanderbilt, David and Xiao, Di},
  journal = {Phys. Rev. B},
  volume = {97},
  issue = {13},
  pages = {134423},
  numpages = {8},
  year = {2018},
  month = {Apr},
  publisher = {American Physical Society},
  doi = {10.1103/PhysRevB.97.134423},
  url = {https://link.aps.org/doi/10.1103/PhysRevB.97.134423}
}

@article{sato2026,
  title={{Quantum theory of magnetic octupole in periodic crystals and application to d-wave altermagnets}},
  author={Sato, Takumi and Hayami, Satoru},
  journal={npj Quantum Mater.},
  volume={11},
  pages={32},
  year={2026},
  doi = {10.1038/s41535-026-00865-9},
  url = {https://doi.org/10.1038/s41535-026-00865-9}
}

@article{gao2018orbital,
  title={{Orbital magnetic quadrupole moment and nonlinear anomalous thermoelectric transport}},
  author = {Gao, Yang and Xiao, Di},
  journal = {Phys. Rev. B},
  volume = {98},
  issue = {6},
  pages = {060402},
  numpages = {5},
  year = {2018},
  month = {Aug},
  publisher = {American Physical Society},
  doi = {10.1103/PhysRevB.98.060402},
  url = {https://link.aps.org/doi/10.1103/PhysRevB.98.060402}
}

@article{shi2007quantum,
  title={{Quantum Theory of Orbital Magnetization and Its Generalization to Interacting Systems}},
  author = {Shi, Junren and Vignale, G. and Xiao, Di and Niu, Qian},
  journal = {Phys. Rev. Lett.},
  volume = {99},
  issue = {19},
  pages = {197202},
  numpages = {4},
  year = {2007},
  month = {Nov},
  publisher = {American Physical Society},
  doi = {10.1103/PhysRevLett.99.197202},
  url = {https://link.aps.org/doi/10.1103/PhysRevLett.99.197202}
}

@article{katsura2005spin,
  title={{Spin Current and Magnetoelectric Effect in Noncollinear Magnets}},
  author = {Katsura, Hosho and Nagaosa, Naoto and Balatsky, Alexander V.},
  journal = {Phys. Rev. Lett.},
  volume = {95},
  issue = {5},
  pages = {057205},
  numpages = {4},
  year = {2005},
  month = {Jul},
  publisher = {American Physical Society},
  doi = {10.1103/PhysRevLett.95.057205},
  url = {https://link.aps.org/doi/10.1103/PhysRevLett.95.057205}
}

@article{vsmejkal2022beyond,
  title={{Beyond Conventional Ferromagnetism and Antiferromagnetism: A Phase with Nonrelativistic Spin and Crystal Rotation Symmetry}},
  author = {\ifmmode \check{S}\else \v{S}\fi{}mejkal, Libor and Sinova, Jairo and Jungwirth, Tomas},
  journal = {Phys. Rev. X},
  volume = {12},
  issue = {3},
  pages = {031042},
  numpages = {16},
  year = {2022},
  month = {Sep},
  publisher = {American Physical Society},
  doi = {10.1103/PhysRevX.12.031042},
  url = {https://link.aps.org/doi/10.1103/PhysRevX.12.031042}
}

@article{PhysRev.95.1154,
  title={{Hall Effect in Ferromagnetics}},
  author = {Karplus, Robert and Luttinger, J. M.},
  journal = {Phys. Rev.},
  volume = {95},
  issue = {5},
  pages = {1154--1160},
  numpages = {0},
  year = {1954},
  month = {Sep},
  publisher = {American Physical Society},
  doi = {10.1103/PhysRev.95.1154},
  url = {https://link.aps.org/doi/10.1103/PhysRev.95.1154}
}

@article{SMIT195839,
title={{The spontaneous hall effect in ferromagnetics II}},
journal = {Physica},
volume = {24},
number = {1},
pages = {39-51},
year = {1958},
issn = {0031-8914},
doi = {https://doi.org/10.1016/S0031-8914(58)93541-9},
url = {https://www.sciencedirect.com/science/article/pii/S0031891458935419},
author = {J. Smit}
}

@article{PhysRev.160.421,
  title={{Contributions to the Theory of the Anomalous Hall Effect in Ferro- and Antiferromagnetic Materials}},
  author = {Maranzana, F. E.},
  journal = {Phys. Rev.},
  volume = {160},
  issue = {2},
  pages = {421--429},
  numpages = {0},
  year = {1967},
  month = {Aug},
  publisher = {American Physical Society},
  doi = {10.1103/PhysRev.160.421},
  url = {https://link.aps.org/doi/10.1103/PhysRev.160.421}
}

@article{berger1970slide,
  title={{Side-Jump Mechanism for the Hall Effect of Ferromagnets}},
  author = {Berger, L.},
  journal = {Phys. Rev. B},
  volume = {2},
  issue = {11},
  pages = {4559--4566},
  numpages = {0},
  year = {1970},
  month = {Dec},
  publisher = {American Physical Society},
  doi = {10.1103/PhysRevB.2.4559},
  url = {https://link.aps.org/doi/10.1103/PhysRevB.2.4559}
}

@article{nozieres1973simple,
  title={A simple theory of the anomalous Hall effect in semiconductors},
  author={Nozi{\`e}res, Ph and Lewiner, CJJPF},
  journal={Journal de Physique},
  volume={34},
  number={10},
  pages={901--915},
  year={1973},
  publisher={Soci{\'e}t{\'e} Fran{\c{c}}aise de Physique},
  doi={10.1051/jphys:019730034010090100},
  url={https://doi.org/10.1051/jphys:019730034010090100}
}

@article{ye1999berry,
  title={{Berry Phase Theory of the Anomalous Hall Effect: Application to Colossal Magnetoresistance Manganites}},
  author = {Ye, Jinwu and Kim, Yong Baek and Millis, A. J. and Shraiman, B. I. and Majumdar, P. and Te\ifmmode \check{s}\else \v{s}\fi{}anovi\ifmmode \acute{c}\else \'{c}\fi{}, Z.},
  journal = {Phys. Rev. Lett.},
  volume = {83},
  issue = {18},
  pages = {3737--3740},
  numpages = {0},
  year = {1999},
  month = {Nov},
  publisher = {American Physical Society},
  doi = {10.1103/PhysRevLett.83.3737},
  url = {https://link.aps.org/doi/10.1103/PhysRevLett.83.3737}
}

@article{jungwirth2002,
  title={{Anomalous Hall Effect in Ferromagnetic Semiconductors}},
  author = {Jungwirth, T. and Niu, Qian and MacDonald, A. H.},
  journal = {Phys. Rev. Lett.},
  volume = {88},
  issue = {20},
  pages = {207208},
  numpages = {4},
  year = {2002},
  month = {May},
  publisher = {American Physical Society},
  doi = {10.1103/PhysRevLett.88.207208},
  url = {https://link.aps.org/doi/10.1103/PhysRevLett.88.207208}
}

@article{nagaosa2010anomalous,
  title={{Anomalous Hall effect}},
  author = {Nagaosa, Naoto and Sinova, Jairo and Onoda, Shigeki and MacDonald, A. H. and Ong, N. P.},
  journal = {Rev. Mod. Phys.},
  volume = {82},
  issue = {2},
  pages = {1539--1592},
  numpages = {0},
  year = {2010},
  month = {May},
  publisher = {American Physical Society},
  doi = {10.1103/RevModPhys.82.1539},
  url = {https://link.aps.org/doi/10.1103/RevModPhys.82.1539}
}

@article{gosalbez2015chiral,
  title={{Chiral degeneracies and Fermi-surface Chern numbers in bcc Fe}},
  author = {Gos\'albez-Mart\'{\i}nez, Daniel and Souza, Ivo and Vanderbilt, David},
  journal = {Phys. Rev. B},
  volume = {92},
  issue = {8},
  pages = {085138},
  numpages = {22},
  year = {2015},
  month = {Aug},
  publisher = {American Physical Society},
  doi = {10.1103/PhysRevB.92.085138},
  url = {https://link.aps.org/doi/10.1103/PhysRevB.92.085138}
}

@article{hayami2023time,
  title={{Time-reversal switching responses in antiferromagnets}},
  author = {Hayami, Satoru and Kusunose, Hiroaki},
  journal = {Phys. Rev. B},
  volume = {108},
  issue = {14},
  pages = {L140409},
  numpages = {6},
  year = {2023},
  month = {Oct},
  publisher = {American Physical Society},
  doi = {10.1103/PhysRevB.108.L140409},
  url = {https://link.aps.org/doi/10.1103/PhysRevB.108.L140409}
}

@article{kato2026electric,
  title={{Electric-field-induced magnetic toroidal moment and nonlinear magnetoelectric effect in antiferromagnetic olivines}},
  author = {Kato, Yasuyuki and Hayashida, Takeshi and Matsumoto, Koei and Kimura, Tsuyoshi and Motome, Yukitoshi},
  journal = {Phys. Rev. B},
  volume = {113},
  issue = {9},
  pages = {094411},
  numpages = {10},
  year = {2026},
  month = {Mar},
  publisher = {American Physical Society},
  doi = {10.1103/7m1w-rpyn},
  url = {https://link.aps.org/doi/10.1103/7m1w-rpyn}
}

@article{hayashida2025electric,
author = {Hayashida, Takeshi and Matsumoto, Koei and Kimura, Tsuyoshi},
title={{Electric Field-Induced Nonreciprocal Directional Dichroism in a Time-Reversal-Odd Antiferromagnet}},
journal = {Advanced Materials},
volume = {37},
number = {9},
pages = {2414876},
doi = {https://doi.org/10.1002/adma.202414876},
url = {https://advanced.onlinelibrary.wiley.com/doi/abs/10.1002/adma.202414876},
year = {2025}
}

@article{hayami2024analysis,
author = {Hayami, Satoru and Yambe, Ryota and Kusunose, Hiroaki},
title={{Analysis of Photo-Induced Chirality and Magnetic Toroidal Moment Based on Floquet Formalism}},
journal = {J. Phys. Soc. Jpn.},
volume = {93},
number = {4},
pages = {043702},
year = {2024},
doi = {10.7566/JPSJ.93.043702},
URL = {https://doi.org/10.7566/JPSJ.93.043702},
}

@article{garlea2008magnetic,
  title = {{Magnetic and Orbital Ordering in the Spinel ${\mathrm{MnV}}_{2}{\mathrm{O}}_{4}$}},
  author = {Garlea, V. O. and Jin, R. and Mandrus, D. and Roessli, B. and Huang, Q. and Miller, M. and Schultz, A. J. and Nagler, S. E.},
  journal = {Phys. Rev. Lett.},
  volume = {100},
  issue = {6},
  pages = {066404},
  numpages = {4},
  year = {2008},
  month = {Feb},
  publisher = {American Physical Society},
  doi = {10.1103/PhysRevLett.100.066404},
  url = {https://link.aps.org/doi/10.1103/PhysRevLett.100.066404}
}

@article{KNIGHT2020155935,
title={{Nuclear and magnetic structures of KMnF3 perovskite in the temperature interval 10 K–105 K}},
journal = {J. Alloys Compd.},
volume = {842},
pages = {155935},
year = {2020},
issn = {0925-8388},
doi = {https://doi.org/10.1016/j.jallcom.2020.155935},
url = {https://www.sciencedirect.com/science/article/pii/S0925838820322994},
author = {Kevin S. Knight and Dmitry D. Khalyavin and Pascal Manuel and Craig L. Bull and Paul McIntyre}
}

@article{garca1991complex,
  title={{Complex magnetic structures of the rare-earth cuprates ${\mathit{R}}_{2}$${\mathrm{Cu}}_{2}$${\mathrm{O}}_{5}$ (R=Y,Ho,Er,Yb,Tm)}},
  author = {Garca-Muoz, J. L. and Rodrguez-Carvajal, J. and Obradors, X. and Vallet-Reg, M. and Gonz\'alez-Calbet, J. and Parras, M.},
  journal = {Phys. Rev. B},
  volume = {44},
  issue = {9},
  pages = {4716--4719},
  numpages = {0},
  year = {1991},
  month = {Sep},
  publisher = {American Physical Society},
  doi = {10.1103/PhysRevB.44.4716},
  url = {https://link.aps.org/doi/10.1103/PhysRevB.44.4716}
}

@article{morosan2008structure,
  title={{Structure and magnetic properties of the ${\text{Ho}}_{2}{\text{Ge}}_{2}{\text{O}}_{7}$ pyrogermanate}},
  author = {Morosan, E. and Fleitman, J. A. and Huang, Q. and Lynn, J. W. and Chen, Y. and Ke, X. and Dahlberg, M. L. and Schiffer, P. and Craley, C. R. and Cava, R. J.},
  journal = {Phys. Rev. B},
  volume = {77},
  issue = {22},
  pages = {224423},
  numpages = {7},
  year = {2008},
  month = {Jun},
  publisher = {American Physical Society},
  doi = {10.1103/PhysRevB.77.224423},
  url = {https://link.aps.org/doi/10.1103/PhysRevB.77.224423}
}

@article{li2014magnetic,
author = {Li, Man-Rong and Retuerto, Maria and Walker, David and Sarkar, Tapati and Stephens, Peter W. and Mukherjee, Swarnakamal and Dasgupta, Tanusri Saha and Hodges, Jason P. and Croft, Mark and Grams, Christoph P. and Hemberger, Joachim and Sánchez-Benítez, Javier and Huq, Ashfia and Saouma, Felix O. and Jang, Joon I. and Greenblatt, Martha},
title={{Magnetic-Structure-Stabilized Polarization in an Above-Room-Temperature Ferrimagnet}},
journal = {Angew. Chem. Int. Ed.},
volume = {53},
number = {40},
pages = {10774-10778},
doi = {https://doi.org/10.1002/anie.201406180},
url = {https://onlinelibrary.wiley.com/doi/abs/10.1002/anie.201406180},
year = {2014}
}

@article{sundaram1999wave,
  title={{Wave-packet dynamics in slowly perturbed crystals: Gradient corrections and Berry-phase effects}},
  author = {Sundaram, Ganesh and Niu, Qian},
  journal = {Phys. Rev. B},
  volume = {59},
  issue = {23},
  pages = {14915--14925},
  numpages = {0},
  year = {1999},
  month = {Jun},
  publisher = {American Physical Society},
  doi = {10.1103/PhysRevB.59.14915},
  url = {https://link.aps.org/doi/10.1103/PhysRevB.59.14915}
}

@article{ma2010abelian,
  title={{Abelian and non-Abelian quantum geometric tensor}},
  author = {Ma, Yu-Quan and Chen, Shu and Fan, Heng and Liu, Wu-Ming},
  journal = {Phys. Rev. B},
  volume = {81},
  issue = {24},
  pages = {245129},
  numpages = {5},
  year = {2010},
  month = {Jun},
  publisher = {American Physical Society},
  doi = {10.1103/PhysRevB.81.245129},
  url = {https://link.aps.org/doi/10.1103/PhysRevB.81.245129}
}

@article{kang2025measurements,
  title={{Measurements of the quantum geometric tensor in solids}},
  author={Kang, Mingu and Kim, Sunje and Qian, Yuting and Neves, Paul M and Ye, Linda and Jung, Junseo and Puntel, Denny and Mazzola, Federico and Fang, Shiang and Jozwiak, Chris and others},
  journal={Nature Physics},
  volume={21},
  number={1},
  pages={110--117},
  year={2025},
  publisher={Nature Publishing Group UK London},
  doi = {10.1038/s41567-024-02678-8},
  url = {https://doi.org/10.1038/s41567-024-02678-8}
}

@misc{sato2026orbitalmagneticoctupolecrystalline,
      title={Orbital magnetic octupole in crystalline solids and characterization of orbital altermagnetism}, 
      author={Takumi Sato and Satoru Hayami},
      year={2026},
      eprint={2512.24269},
      archivePrefix={arXiv},
      primaryClass={cond-mat.mes-hall},
      url={https://arxiv.org/abs/2512.24269}, 
}

@article{chang1995berry,
  title={Berry phase, hyperorbits, and the Hofstadter spectrum},
  author={Chang, Ming-Che and Niu, Qian},
  journal={Phys. Rev. Lett.},
  volume={75},
  number={7},
  pages={1348},
  year={1995},
  publisher={APS},
  doi = {10.1103/PhysRevLett.75.1348},
  url = {https://doi.org/10.1103/PhysRevLett.75.1348}
}

@article{naka2020anomalous,
  title = {Anomalous Hall effect in $\ensuremath{\kappa}$-type organic antiferromagnets},
  author = {Naka, Makoto and Hayami, Satoru and Kusunose, Hiroaki and Yanagi, Yuki and Motome, Yukitoshi and Seo, Hitoshi},
  journal = {Phys. Rev. B},
  volume = {102},
  issue = {7},
  pages = {075112},
  numpages = {11},
  year = {2020},
  month = {Aug},
  publisher = {American Physical Society},
  doi = {10.1103/PhysRevB.102.075112},
  url = {https://link.aps.org/doi/10.1103/PhysRevB.102.075112}
}

@article{naka2022anomalous,
  title = {Anomalous Hall effect in antiferromagnetic perovskites},
  author = {Naka, Makoto and Motome, Yukitoshi and Seo, Hitoshi},
  journal = {Phys. Rev. B},
  volume = {106},
  issue = {19},
  pages = {195149},
  numpages = {11},
  year = {2022},
  month = {Nov},
  publisher = {American Physical Society},
  doi = {10.1103/PhysRevB.106.195149},
  url = {https://link.aps.org/doi/10.1103/PhysRevB.106.195149}
}

\end{document}